\documentclass[11pt]{article}

\usepackage{authblk,blindtext}
\usepackage[square,numbers]{natbib}
\usepackage{amsmath,amssymb,amsthm}
\usepackage{dsfont}
\usepackage{mathrsfs}
\usepackage{stmaryrd}
\usepackage{float}
\usepackage{subfigure}
\usepackage[margin=1in]{geometry}

\usepackage{graphicx}
\usepackage{epstopdf}

\usepackage{xcolor}
\usepackage{tabu}
\usepackage{multicol}
\usepackage{lipsum}	
\usepackage{caption}
\usepackage{longtable}
\usepackage{booktabs}

\DeclareMathAlphabet{\mathpzc}{OT1}{pzc}{m}{it}

\DeclareMathOperator*{\argmin}{arg\,min}

\newtheorem{remark}{Remark}

\definecolor{orange-red}{rgb}{1.0, 0.5, 0.0}

\newcommand{\dd}{\,\text{d}}

\newcommand{\ii}{\text{i}}

\newcommand{\Lambdak}{\Lambda_{\kappa}}

\newcommand{\obs}{_{\text{obs}}}

\renewcommand{\Re}{\text{Re}}
\renewcommand{\Im}{\text{Im}}

\newcommand{\hh}{\hspace*{0.7pt}}

\newcommand{\bp}{\textbf{\text{p}}}

\begin{document}

\title{Ultrasonic sensing of the mechanical fingerprint of reactive transport in rock}

\author[1]{Ali Aminzadeh}
\author[1]{Prasanna Salasiya}
\author[1]{Joseph F. Labuz}
\author[2]{Mohammad Nooraiepour}
\author[1]{Bojan B. Guzina\thanks{Corresponding author: guzin001@umn.edu}}

\affil[1]{Department of Civil, Environmental, and Geo- Engineering, University of Minnesota, Minneapolis, MN, US }
\affil[2]{Department of Geosciences, University of Oslo, Oslo, Norway}

\date{\today}

\maketitle

\begin{abstract} 
\noindent Mineral carbon storage in rock formations has gained significant interest in recent years. In principle, changes in mechanical rock properties driven by carbon mineralization could be quantified using seismic methods, opening the door toward field monitoring of (the progress of) carbon storage. However, these changes may vary spatially within a rock mass when reactive transport occurs. In this vein, full-field ultrasonic characterization of reacted specimens can help shed light on the process. We use a 3D Scanning Laser Doppler Vibrometer to perform full-field monitoring of one-dimensional (1D) ultrasonic waves in rod-shaped sandstone specimens exposed to NaCl-rich fluid. Our initial experiments were conducted on intact sandstone specimens with high aspect ratio ($\text{length/diameter}\simeq15$) to cater for 1D axial wave propagation. To investigate the evolution of the Young's modulus and attenuation of rock  due to reactive  transport, we exposed the specimens to an under-saturated NaCl solution, achieving supersaturation - and so mineralization -- through evaporation. The upward movement of the fluid, supplied at the  bottom of each specimen, was achieved through capillary action. We deploy an elastography-type approach to back-analysis, known as modified-error-in-constitutive-relation (MECR) approach, to expose the spatially-heterogeneous evolution of mechanical rock properties due to reactive transport. Our results consistently demonstrate (i) degradation of the Young's modulus, and (ii) increase in attenuation due to mineralization. To better understand the root causes of these changes, we made use of the X-ray micro-computed tomography ($\mu$CT) and scanning electron microscopy (SEM) of selected cross-sections. The grain-scale information suggests that  microcracking (resp. pore filling) is predominantly (resp. partly) responsible for the observed macroscopic changes. 
\end{abstract}

\section{Introduction} \label{sec:introduction}

\noindent Salt weathering in porous materials is an important phenomenon that occurs when salts mineralize within voids. Salt weathering occurs in a variety of materials like soil~\citep{Chhabra2022}, rock~\citep{Hall2001}, concrete~\citep{Pel2003}, and wood~\citep{Mi2020}. Numerous studies have considered the fundamental processes behind salt weathering \citep[e.g.][]{Taber1916, Steiger2005, Doehne2010, shahidzadeh2024crystallization}, as well as its effects on mechanical properties such as ultrasonic wave velocity~\cite{Khodabandeh2022}. For instance, a decrease of Young’s modulus due to salt mineralization in a sandstone has been documented~\cite{nooraiepour2025potential}.

Another mineralization process of great interest involves CO2 storage in geological formations~\cite{Ringrose2019}. One such approach is the use of saline aquifers due to their high storage capacity and long lifetime~\cite{Bruant2002}, where the interplay between carbon storage in the porous matrix and changes in mechanical properties must be accurately measured. For example, a decrease in stiffness and an increase in porosity for Entrada sandstone due to CO2-charged brine has been reported~\cite{Espinoza2018}, as was the deterioration of Hawkesbury sandstone by CO2 injection and pore fluid salinity effects~\cite{Rathnaweera2015}.

Injection of dry or undersaturated CO2 (with respect to water) induces drying of brine pore fluids, thereby triggering salt mineralization~\citep{Masoudi2021,nooraiepour2018effect,dkabrowski2025surface}. Further, it has been shown that salt mineralization in porous rocks can vary as a result of the type of salt, apart from rock properties~\cite{Rodriguez1999}. In particular, sodium sulfate generates subflorescence (inside the porous material), which is more damaging, while sodium chloride forms efflorescence (on the surface of the porous material) or propagates by homogeneously filling the tiniest pores within the porous material. In~\cite{Rodriguez1999} it was also observed that lower relative humidity led to faster solution evaporation, resulting in higher levels of supersaturation, which in turn caused greater damage to the rock.

Changes in mechanical rock properties due to mineralization can in principle be sensed by seismic methods. However, most relevant studies report only overall changes in the phase velocity of seismic (or ultrasonic) waves propagated through a material, which does not provide insight into spatially-dependent matrix alterations within the rock mass. To overcome this limitation, various approaches to the back-analysis covered by the umbrella term of \emph{elastography}~\cite{greenleaf2003, parker2005}, commonly used in medical diagnosis~\cite{sigrist2017, mariappan2010}, can be employed to characterize the rock properties in situations where full-field (``interior'') observations of the propagated wavefields are available. For instance, the authors in~\cite{Pourahmadian2018} used direct algebraic inversion to reconstruct the distribution of heterogeneous Young's modulus in a granite slab, where the interior (in-plane) ultrasonic waveforms propagated through the specimen were captured by scanning laser Doppler vibrometry. When seeking to expose the mechanical fingerprint of reactive transport in rock, however, it is important to recognize that the reaction-driven pore filling and microcracking could in general alter both dispersion and attenuation of seismic or ultrasonic waves. To incorporate such mechanisms into our characterization framework, we model both intact and reacted rock as a \emph{heterogeneous viscoelastic solid}. Under such hypothesis, an extension in~\cite{bonnet2024} of the modified-error-in-constitutive (MECR) approach~\cite{lad:ned:rey:94,allix:05} to elastography, formulated to account for material dissipation (and so wave dispersion) effects, can be applied to enable the reconstruction of viscoelastic rock properties from the full-field observations of  ultrasonic waves captured by a 3D Scanning Laser Doppler Vibrometer (SLDV).

In this study we generate, and investigate the mechanical fingerprint of, \emph{one-dimensional} reactive transport by (i) preparing rod-like sandstone specimens with the aspect ratio of 15;  (ii) submerging one end of the rods in sodium chloride (NaCl); (iii) allowing for the reactive transport and mineralizaton to develop respectively by capillary action and evaporation; (iv) capturing the full-field ultrasonic waveforms propagating along the length of a rod by a 3D laser Doppler vibrometer; and (v) using the MECR methodology to interpret the data. To contrast the so-exposed \emph{macroscopic} variation  in mechanical properties against the \emph{microscopic} (pore-scale) changes, we also acquired the micro-computed X-ray tomography ($\mu$CT) and scanning electron microscopy (SEM) images from the bottom, middle, and top sections of each specimen before and after mineralization. Our samples were exposed to NaCl solution for a shorter period of time (two weeks) compared to previous related studies \citep[e.g.][]{Rodriguez1999, Scherer2004} to prevent surface peeling, which is known to occur after prolonged exposure to a source of reactive fluid. Specifically, our goal was to expose the mechanical trace of mineralization (via  ultrasonic measurements) while preserving the integrity of the surface of the specimen, i.e., when there is no visible damage to the specimen. 

The paper is organized as follows. Section~\ref{sec:method} describes specimen preparation, nominal material properties, test procedure, ultrasonic excitation, and signal processing techniques. In Section~\ref{sec:results} we present the results of experiments and back-analysis, consistently demonstrating a drop in the Young's modulus and a marked increase in attenuation coefficient after mineralization. Section~\ref{sec:discussion} and Section~\ref{sec:conclusions} are devoted respectively to the discussion and summary of our findings.

\section{Methodology} \label{sec:method}

\subsection{Specimen preparation} \label{sub:preparation} 

\begin{figure}[H]
\center
\includegraphics[height=9 cm]{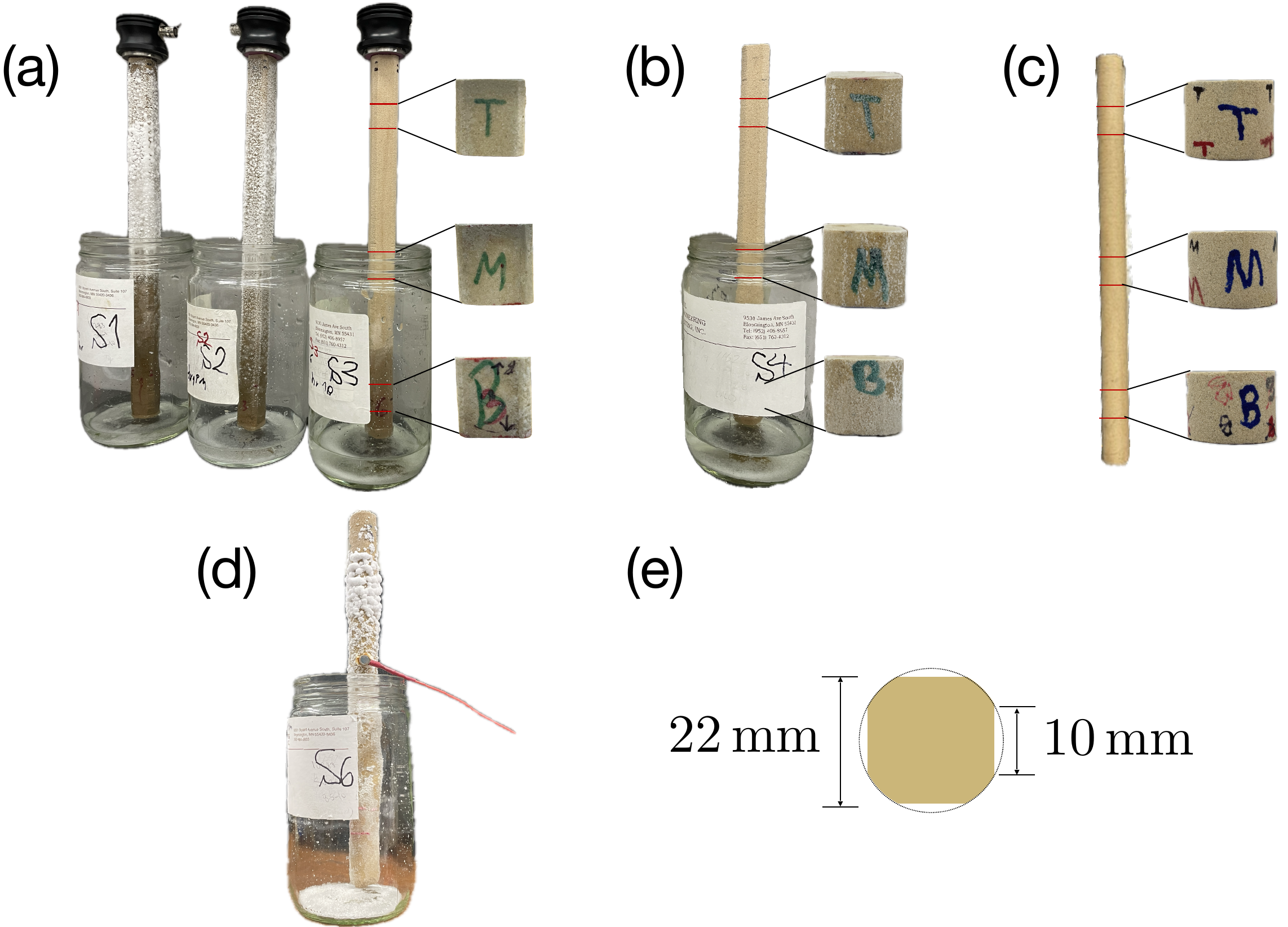}
\caption{Rod-like specimens of Dunnville sandstone: (a) S$1$-S$3$, (b) S$4$, (c) S$5$,  (d) S$6$, and (e)~geometry of their cross section featuring four ``sensing strips''.}
\label{fig:specimens}
\end{figure}

Six sandstone rods were prepared by coring a large block of Dunnville sandstone (Wisconsin, US) perpendicular to its bedding plane to obtain slender prismatic specimens with length $\ell=300\,\text{mm}$ and diameter $d=22\,\text{mm}$ ($\ell/d\simeq 15$). Then, four sides of each rod were ground as in Fig.~\ref{fig:specimens}(e) to obtain $10\,\text{mm} \times 300\,\text{mm}$ flat ``sensing strips'' to be scanned by SLDV during ultrasonic testing. With reference to Fig.~\ref{fig:specimens}, three specimens (S1 through~S3) were used for the ultrasonic experiments; two additional specimens (S4 and S5) were used, together with S$3$, for the $\mu$CT and SEM scanning; and one specimen (S$6$) was dedicated to monitoring the acoustic emission activity due to reaction-driven cracking. 

\subsection{Material properties} \label{sub:properties} 

Mechanical properties of Dunnville sandstone were obtained through ultrasonic testing on the large block before coring, by measuring the $P$- and $S$-wave velocities (perpendicular to the bedding plane) using  GCTS ULT-200 transducers operating at 1$\,$MHz. A shear coupling gel was used to facilitate the transmission of ultrasonic energy between the transmitter/receiver and the surface of the specimen. The resulting $P$-wave, $S$-wave and longitudinal ``rod'' velocities ($c_{P}$, $c_{S}$, and $c_{L}$), mass density~$\rho$, Young's and shear moduli ($E$ and~$G$), and Poisson's ratio~$\nu$ of Dunville sandstone are summarized in Table \ref{tab:properties}. Comparable properties have also been reported in other studies~\citep[e.g.][]{Tarokh2022}. Notably, Dunville sandstone exhibits high porosity ($n\simeq30\%$) and homogeneity, with a mineralogical composition of 90--95\% quartz.
\begin{table}[H]
\centering
\caption{Nominal elastic properties of Dunnville sandstone obtained from the ultrasonic $P$- and $S$-wave measurements at 1$\,$MHz}.
\begin{tabu}{ccccccc} \tabucline[1pt]{-} \addlinespace[2pt]
$c_{P}$ & $c_{S}$ & $c_{L}$ & $\rho$     &  $E$   &   $G$   & $\nu$ \\ \relax
[km/s]  &  [km/s] &  [km/s] & [g/cm$^3$] & [GPa]  &  [GPa]  &  \\ 
\addlinespace[2pt] \tabucline[1pt]{-} \addlinespace[2pt]
1.86 & 1.20 & 1.81 & 1.84 & 6.2  & 2.7 & 0.15 \\  
\tabucline[1pt]{-}
\end{tabu}
\label{tab:properties}
\end{table}

\subsection{Testing procedure} \label{sub:procedure} 

\begin{figure}[]
\center
\includegraphics[height=7 cm]{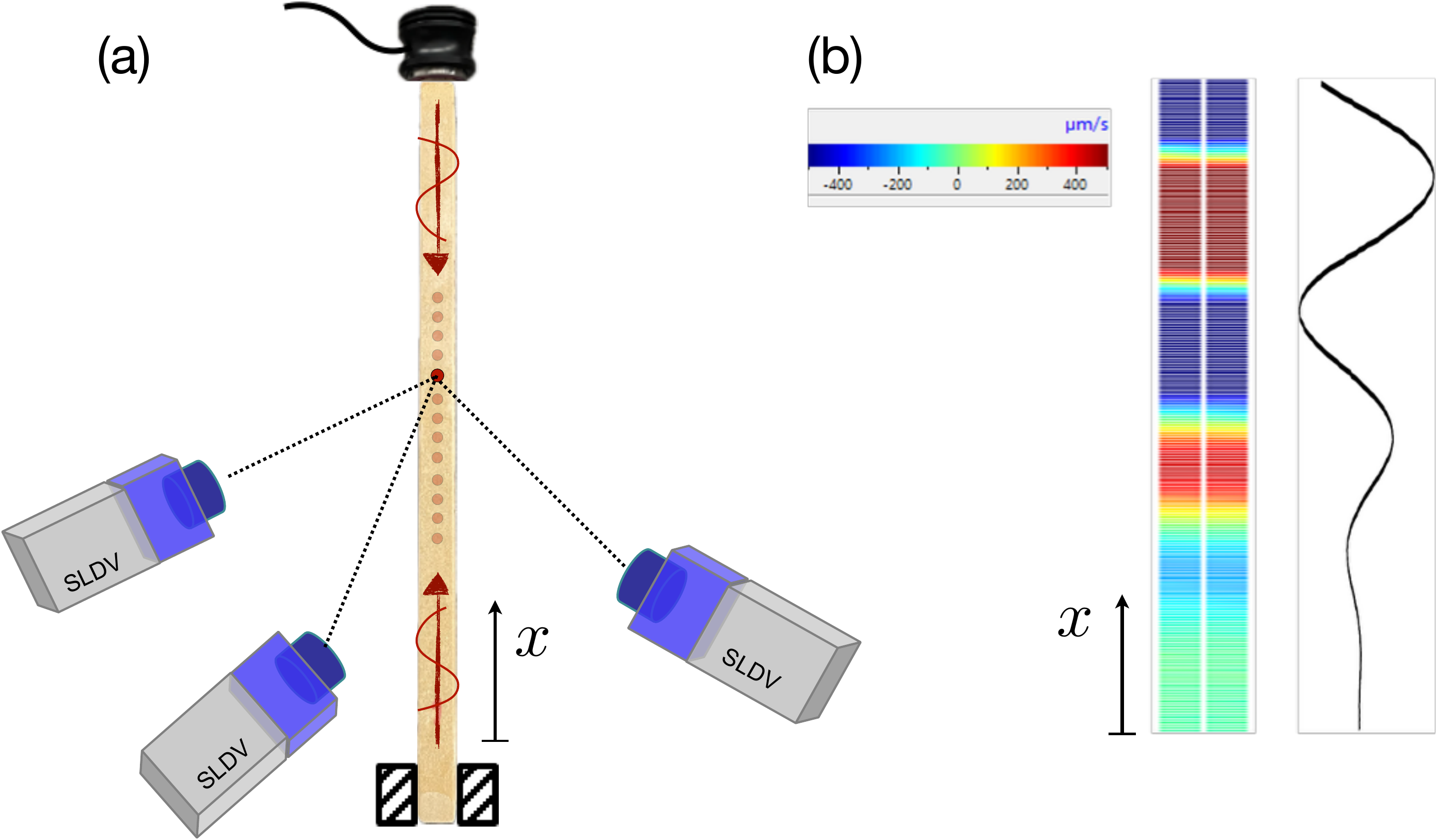}
\caption{SLDV testing: (a) Schematics of a line scan, and (b) example snapshot of the recorded axial velocity profile $v^{\obs}(x,t)$ at $t=0.15\,$ms.}
\label{fig:laser}
\end{figure}

In each experiment, sandstone specimen was fixed at the bottom and excited at the top by a transient wavelet emanating from an ultrasonic transducer (see Sec.~\ref{sub:wavelet} for details). To capture the so-generated axial motion along the length of the rod, we deployed a 3D Scanning Laser Doppler Vibrometer (SLDV) system PSV-400 by Polytec, Inc. featuring 0.1$\;$mm spatial resolution, down to O($n$m) displacement sensitivity, and operating  frequency range DC--1$\;$MHz. In this way, full-field waveforms of axial motion through the rod-shaped specimens (S1--S3) were scanned with 1$\;$mm axial resolution as shown Fig.~\ref{fig:laser}(a).
Over any  given (10$\,$mm-wide) sensing strip featured in Fig.~\ref{fig:specimens}(e), we performed three 250$\,$-mm long line scans (2.5$\,$mm apart) starting 10 mm from the top of the specimen. The axial particle velocity as a function of time and space, $v^{\obs}(x,t)$ -- an example of which is shown in Fig.~\ref{fig:laser}(b) -- is then computed by averaging the six line scans (three scans each taken along the ``front'' and ``back'' sensing strip). To minimize the adverse effect of random measurement errors, the particle velocity at each scan point is computed by averaging over 60 realizations of the ultrasonic experiment. 

\emph{\bfseries Tests on intact specimens: Series~I.} The first batch of ultrasonic experiments, hereon referred to as Series~I, were performed on intact specimens. Prior to performing the reference scans,  intact samples were oven-dried at a temperature of $50^{\circ}C$ for five days, with the drying temperature kept low to minimize the potential for thermal cracking.

\emph{\bfseries Tests on reacted specimens: Series~II.} To investigate the changes in mechanical (elastic and dissipative) properties of rock due to exposure to reactive fluid transport,  intact specimens S1-S3 were exposed from the bottom as in Fig. \ref{fig:specimens}(a) to a solution that was initially undersaturated with respect to NaCl, which allowed for (i) the solute to move upward by capillary action and (ii) supersaturation of the pore fluid to develop gradually by evaporation \citep{Scherer2004}. In each experiment, the solution supply consisted of 71.8$\,$g of sodium chloride (NaCl) fully dissolved in 200$\,$ml of deionized (DI) water. The bottom of each specimen was kept in the solution for~14 days at room temperature. After the two-week reaction period, the outer surface of each specimen was cleaned of softly attached salt deposits (see~Fig.~\ref{fig:specimens}(d) for instance). Then, the specimens were oven-dried for five days at $50^{\circ}C$ and tested dynamically by SLDV (Series~II). 

\emph{\bfseries Tests on re-reacted specimens: Series~III.} After completing the Series~II of ultrasonic experiments, specimens S1-S3 were subjected to an additional reactive fluid transport by: (i) placing the bottom of each specimen in a solution for one week, with the solution containing the same amount of salt and water as in the original procedure; (ii) leaving the specimen outside of the solution for one week (at a room temperature)
to allow the halite crystals to dry and harden around the rod; and (iii) placing the specimen back in a solution for one more week. The so-reacted specimens were then placed in an oven for five days, after which period the outer surface of each specimen was cleaned of the hardened halite minerals. 

The preparation of specimens S4 and S5 adhered to the procedure described above for Series~II and Series~I, respectively. The microstructural ($\mu$-CT and SEM) analysis of specimens S3, S4, and S5 corresponding respectively to Series~III, Series~II, and Series~I was performed on representative (20$\,$mm-thick) slices taken near the bottom, middle, and top of each specimen. To better understand the mechanical fingerprint of reactive fluid transport, an additional microstructural characterization of specimen S4 (bottom slice) was performed after immersing it in DI water for five days and oven-drying it for five days. The idea behind this additional test was to ``wash off'' the accumulated precipitate end so help expose the damaged grain structure. The washing process resulted in the removal of 79$\,$mg (about 1\% of the sample mass) of the NaCl precipitate and potentially crushed grain material. The sample preparation process for the three series of tests is summarized in Table~\ref{tab:test procedure}.
\begin{table}[H]
\centering
\caption{Sample preparation and testing summary for specimens S1-S5.}
\begin{tabu}{cccc} \tabucline[1pt]{-} \addlinespace[2pt]
Test series &  I  & II & III   \\ 
\addlinespace[2pt] \tabucline[1pt]{-} \addlinespace[2pt]
SLDV testing & S1-S3  &  S1-S3    &  S1-S3 \\  
Microstructural imaging & S5  &  S4$^\star$   &  S3 \\  
Reactive transport [days] & 0  &  14    &  14$^\dagger$ \\  
Drying in oven [days]  & 5 &   5   &  5 \\  
\tabucline[1pt]{-}
\end{tabu}
\label{tab:test procedure} \vspace*{3mm}
\begin{minipage}{15cm}
{\small $^\star$Used twice for imaging: (a) after Series~II  preparation procedure, and (b) after Series~III preparation procedure followed by the ``specimen-washing'' process.} \\
{\small $^\dagger$With a one-week ``break'' at room temperature between the two weeks of reactive transport.}
\end{minipage}
\end{table}

\subsection{Monitoring of the AE events} \label{sub:AE2} 

To track the microseismic activity caused by reactive transport, an acoustic emission (AE) sensor was glued near the middle section of specimen~S6 as shown in Fig.~\ref{fig:specimens}(d), and the AE events were counted. The deployed sensor, Nano$30$ by Physical Acoustics,  is characterized by the resonant frequency of 300 kHz and has a sensing bandwidth of 125--750$\,$kHz. The events were recorded by an oscilloscope (Picoscope 5000) with a maximum sampling frequency of 1 GS/s, 16-bit resolution, and a bandwidth of 200 MHz.

\subsection{Dynamic excitation for SLDV testing} \label{sub:wavelet} 

\begin{figure}[H]
\center
\includegraphics[width=0.90\linewidth]{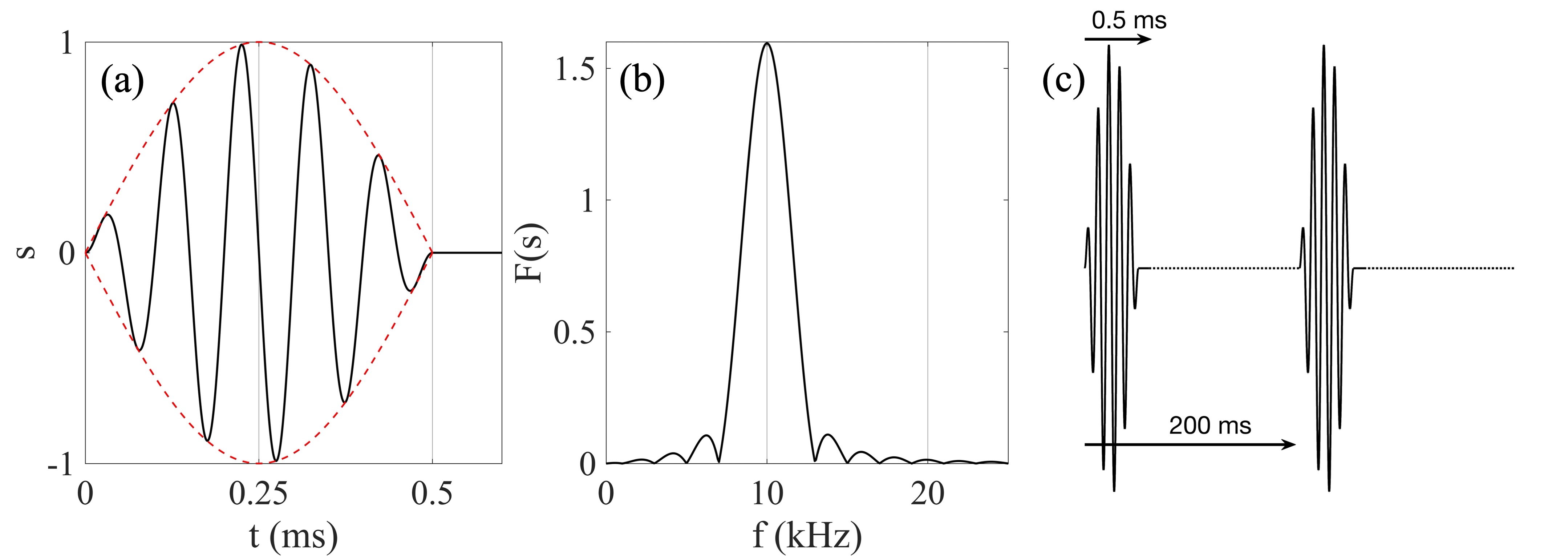}
\caption{Excitation wavelet with a center frequency of $10$ kHz in: (a) time domain, (b) frequency domain, and (c) and its repeated application to the transducer.}
\label{fig:wavelet}
\end{figure}

With reference to Fig.~\ref{fig:laser}(a), transient excitation is applied to the top the specimen by an ultrasonic transducer ($0.5$ MHz Olympus V$101$) affixed with cyanoacrylate glue, while the bottom of the specimen is kept fixed with a heavy vise. During interrogation, specimens were excited by a five-cycle wavelet 
\begin{equation} \label{eq:wavelet}
s(t) = H(t) \; H(5 - f_c\hh t) \; \sin(0.2\pi\hh f_c\hh t) \; \sin(2\pi\hh f_c\hh t), 
\end{equation}
where $H$ and $f_c$ denote respectively the Heaviside step function and center frequency. To help maintain a ``true'' 1D wave motion throughout the specimen, the center frequency is taken as $f\!\in\!\{10,15,20\} \, \text{kHz}$, resulting in the nominal wavelengths $9.1\text{cm}\!<\!\lambda\!<\!18.2 \text{cm}$ that are more than four times greater than the diameter of the rod ($2.2$ cm) and shorter than the length of the specimen ($30$ cm). For completeness, the featured wavelet is illustrated in Fig.~\ref{fig:wavelet}. To both (i) enable trace stacking at a fixed focal point and (ii) allow the SLDV system to scan the entire wavefield (one focal point at a time), the specimen is excited by a periodic application of the wavelet as shown in Fig.~\ref{fig:wavelet}(c). The delay  between successive applications of the wavelet is taken as  200$\,$ms, which allowed for the longitudinal waves in a specimen to die out by material dissipation.

\subsection{Data analysis by the viscoelastic MECR approach} \label{MECR-method}

\begin{figure}[h]
\center
\includegraphics[width = 0.5\textwidth]{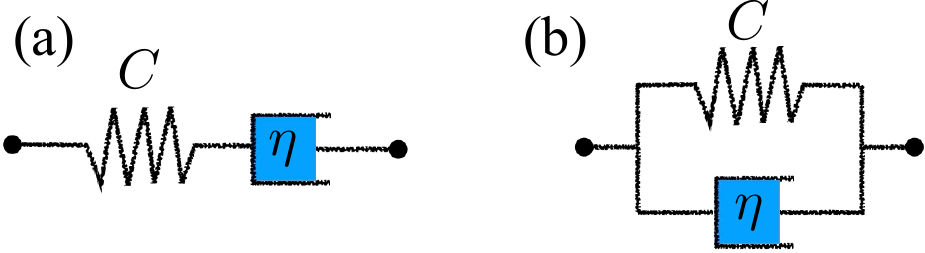}
\caption{Elemental (two-parameter) phenomenological models of the viscoelastic behavior: (a)~Maxwell model, and (b) Kelvin-Voigt model.}
\label{fig:viscomodels}
\end{figure}

The modified-error-in-constitutive-relation (MECR) technique belongs to a class of inverse solutions, frequently used in medical diagnosis~\cite{sigrist2017, mariappan2010}, that are capable of reconstructing the spatially-heterogeneous elastic (and more recently dissipative) properties of a material from the knowledge of an \emph{interior deformation map}, $u^{\obs}$, within a specimen. In this study, $u^{\obs}\!=\!u^{\obs}(x,t)$ is given by the spatiotemporal distribution of longitudinal ultrasonic displacements along the length of a rod; a sensory information that we seek to leverage toward exposing the \emph{macroscopic} mechanical fingerprint of  reaction-driven, \emph{microscopic} changes in rock due to pore filling and microcracking. To this end, we process the sensory data in the frequency domain and model Dunville sandstone as a linear viscoelastic solid endowed with (heterogeneous) complex Young's modulus $\mathpzc{E}$. As shown recently in~\cite{guzina2025constitutive}, the frequency variation of~$\mathpzc{E}$ corresponding to an arbitrary arrangement of ``springs'' and ``dashpots'' (e.g. Maxwell or Kelvin-Voigt model) is given by a rational function  $\mathpzc{E} = \mathpzc{E}(\bp,\omega)$, where $\bp$ is the vector of viscoelastic parameters collecting the featured ``spring'' constants (i.e. elastic moduli) and ``dashpot'' constants (i.e. viscous moduli). Within the MECR framework, elastic and dissipative properties of a viscoelastic solid are then exposed by minimizing a linear combination of the \emph{constitutive mismatch} and \emph{kinematic data mismatch} described next. Letting~$\hat{f}(\omega)$ denote the Fourier transform of a temporal function~$f(t)$, the constitutive mismatch quantifies the discrepancy between (i) the viscoelastic stress field 
\begin{equation} \label{mecr1}
\hat{\sigma}[\hat{u}](x,\omega) \, = \, \mathpzc{E}\hh \frac{\dd \hat{u}}{\dd x}(x,\omega)
\end{equation}
computed for a trial modulus~$\mathpzc{E}(\bp,\omega)$ and trial displacement field $\hat{u}\!=\!\hat{u}(x,\omega)$, and (ii) dynamically-admissible stress field $\hat{\sigma}(x,\omega)$ -- satisfying the balance of linear momentum $\dd\hat\sigma/\dd x +\rho\hh\omega^2\hh \hat{u} = 0$ -- that is computed independently during the minimization process. As in most studies targeting material characterization, the kinematic data misfit quantifies the discrepancy $\hat{u}-\hat{u}^{\obs}$ over the observation window where the (interior) deformation map is being captured. Assuming the Maxwell viscoelastic model shown in Fig.~\ref{fig:viscomodels}(a) (for example) as the reference model for data interpretation, one-dimensional variant of the MECR functional~\cite{bonnet2024} can be explicitly written as 
\begin{equation}  \label{mecr2}
\Lambda_{\kappa}(\hat{u},\hat{\sigma},\bp;\omega) \:= \tfrac{1}{2} \;
\int_{\Omega} \left( \frac{1}{C} + \frac{2\pi}{\omega\hh\eta} \right) \, |\hat{\sigma} - \hat{\sigma}[\hat{u}]|^2 \dd x \;+\; \tfrac{1}{2} \; \kappa\! \int_{\Omega} |\hat{u}-\hat{u}^{\obs}|^2 \dd x,  
\end{equation}
where $\Omega \!=\! (x_1,x_2)$ is the observation window, and $\kappa\!>\!0$ is the weighting coefficient controlling a trade-off between  the constitutive model veracity and data fidelity. This allows to seek the vector of optimal viscoelastic parameters, $\bp^\star$, by solving the nonlinear minimization problem 
\begin{equation} \label{mecr3}
(\hat{u}{}^\star,\hat{\sigma}{}^\star,\bp^\star) ~=~ \argmin_{\hat{u},\hat{\sigma},\bp}  \: \Lambda_{\kappa}(\hat{u},\hat{\sigma},\bp;\omega) \qquad \text{subject to} \qquad 
\frac{\dd \hat{\sigma}}{\dd x} \,+\, \rho\hh\omega^2 \hat{u} \,=\, 0, 
\end{equation}
see~\cite{bonnet2024} in the context of two- and three-dimensional problems. To implement the differential equation constraint in~\eqref{mecr3}, we recall the weak statement of the balance of linear momentum and pursue unconstrained minimization of the Lagrangian 
\begin{equation} \label{mecr4}
\mathcal{L}(\hat{u},\hat{\sigma},\hat{w},\bp;\omega)  ~=~~ 
\Lambda_{\kappa}(\hat{u},\hat{\sigma},\bp;\omega) \,-\, \Re\Big[\int_\Omega \Big(\sigma \hh \frac{\dd\overline{\hat{w}}}{\dd x} - \rho\hh\omega^2 \hat{u}\,\overline{\hat{w}} \Big)\dd x\Big] 
\end{equation}
where the test function $\hat{w} = \hat{w}(x,\omega)$, also termed the adjoint field, simultaneously plays the role of the Lagrange multiplier. The relevant stationarity conditions of~\eqref{mecr4} then lead to a coupled system of equations for the direct field~($\hat{u}$) and adjoint field ($\hat{w}$) over~$\Omega$, driven respectively by the constitutive mismatch $\hat{\sigma} - \hat{\sigma}[\hat{u}]$ and the data mismatch $\hat{u}-\hat{u}{}^{\obs}$ \cite{diaz2015modified,bonnet2024}, that cater for an effective computation of the gradient $\partial\mathcal{L}/\partial\bp$. 

In situations where the sensory data $\hat{u}^{\obs}$ are captured at multiple frequencies $\omega_m$ ($m\!=\!\overline{1,M}$) as in this study, \eqref{mecr3} and~\eqref{mecr4} are readily generalized via the substitution
\begin{equation} \label{mecr5}
\Lambda_{\kappa}(\hat{u},\hat{\sigma},\bp;\omega) ~~\mapsto~~
\sum_{m=1}^M c_m \,\Lambda_{\kappa}(\hat{u}_m,\hat{\sigma}_m,\bp;\omega_m) 
\end{equation}
where $c_m$ are suitable weights, see~\cite{bonnet2024} for details. In the sequel, we conveniently select $c_m = 2\pi/\omega_m$ which gives higher weight to the lower frequencies. Following~\cite{bonnet2024}, we select the penalty parameter~$\kappa$ in~\eqref{mecr2} via the L-curve criterion by (a) plotting (on a log-log scale)  the constitutive mismatch $\mathcal{E}$ against the data mismatch $\mathcal{M}$ at the minimizer of the $\Lambdak$ -- for a range of trial $\kappa$ values, and (b) selecting the optimal value $\kappa=\kappa^\star$ at the “elbow” of the resulting L-curve.

With reference to Fig.~\ref{fig:viscomodels}, we describe the rate-dependent elastic and dissipative properties of Dunnville sandstone, pre- and post-transformation, via the competing Maxwell and Kelvin-Voight viscoelastic models that each feature an Young's modulus~$C$ and viscous modulus~$\eta$ which gives $\bp=(C,\eta)$ in the context of~\eqref{mecr3} and~\eqref{mecr4}. For the two models, we have  
\begin{equation}\label{mod:complex}
\mathpzc{E}(\bp,\omega) = \frac{C\hh\eta\hh \omega}{\ii\hh C + \eta\hh\omega} \quad \text{(Maxwell)} \qquad\text{and}\qquad 
\mathpzc{E}(\bp,\omega)  = C - \ii \hh\eta\hh\omega  \quad \text{(Kelvin-Voigt)}, 
\end{equation}
which assume implicit time dependence $e^{-\ii\hh\omega\hh t}$ stemming from the definition of the Fourier transform. 
Once the vector of optimal viscoelastic parameters~$\bp^\star$ has been identified, we obtain 
\begin{equation}\label{Edalfa}
E_d(\omega)=\Re[\mathpzc{E}(\bp^\star,\omega)], \qquad
\alpha(\omega) = \Im\Big[ \frac{\omega}{\sqrt{\mathpzc{E}(\bp^\star,\omega) / \rho}} \Big]
\end{equation}
for the frequency-dependent \emph{dynamic Young's modulus} $E_d$ (which captures the effects of wave dispersion) and \emph{attenuation coefficient} $\alpha$ in the rock. 

\paragraph{Selection of the viscoelastic model.} 
In the MECR approach, both the (presumed) constitutive model and sensory data are affiliated with a degree of uncertainty~\cite{banerjee2013}. In this vein, our selection of the appropriate viscoelastic model (between the Maxwell and Kelvin-Voigt variants) is made on the basis of ``goodness of fit'' with multi-frequency data $\hat{u}^{\obs}$ captured at $\omega=\omega_m$ $(m=\overline{1,M})$. Within the MECR framework, the goodness of fit is explicitly quantifiable by the value of the cost functional $\Lambda_\kappa$ due to~\eqref{mecr5} at the minimum, where $\bp=\bp^\star$. Applying this criterion to the SLDV data captured at frequencies $\omega_m = 10\pi\hh (1+m)\times 10^3\,$rad/s ($m=
\overline{1,3}$), we obtained $\Lambda_\kappa^\star=0.02$ for the Maxwell model and $\Lambda_\kappa^\star=0.37$ for the Kelvin-Voigt model, which drove our adoption of the Maxwell model. 

\begin{remark}
With reference to~\eqref{Edalfa}, it is important to note that the dynamic Young's modulus $E_d(\omega)$ caters specifically for capturing the dispersion of ultrasonic waves, whose longitudinal wave speed can be written as $c_L(\omega) = \sqrt{E_d(\omega)/\rho}$. In this vein, the Maxwell viscoelastic model -- while not being relevant in terms of static rock deformation -- has been found to closely describe the variation of longitudinal wave speed $c_L(\omega)$ and attenuation $\alpha(\omega)$ encoded in the SLDV data for frequencies between 10$\,$kHz and 20$\,$kHz.
\end{remark}

\subsection{X-ray tomography} \label{sub:uCT} 

To investigate the pore-scale transformation of Dunnville sandstone caused by reactive transport, three 20$\,$mm-thick sections (top, middle and bottom) were cut out from each specimen S$_5$, S$_4$, and  S$_3$ following the SLDV experiments corresponding to Series~I, Series~II, and Series~III, respectively (see Fig. \ref{fig:specimens}). These sections were scanned by $\mu$CT technique using the XT H 225 scanner by Nikon Metrology, Inc. at a voxel resolution of 20 $\mu\,$m.  The acquisition parameters were set at 110 kV and 90 $\mu$A to ensure optimal image clarity and resolution. The projections were reconstructed using CT Pro 3D XT 3.1.11 software (Nikon Metrology, Inc).

Image processing was carried out using Dragonfly 2024.1 software by Comet Technologies and VGSTUDIO MAX 3.4 by Volume Graphics. Initial noise reduction was achieved through image filtering, which enhanced clarity and improved phase contrast within the dataset. Subsequently, a deep learning-assisted image segmentation approach was employed to differentiate the various components within the tomography images accurately. For this task, a U-Net convolutional neural network~\citep{nooraiepour2025three} was used to distinguish individual constituents. The OpenPNM package \citep{gostick2016openpnm} was employed to extract pore and throat characteristics from the segmented multi-component region of interest, enabling quantitative analysis of the pore network. 

\subsection{Scanning electron microscopy} \label{sub:SEM} 

The same nine sections used for $\mu$CT scanning were coated with a 5–10$\,$nm carbon layer using an SPI module carbon coater to facilitate the SEM analysis of their outer surface. A Thermo Fisher Scientific Apreo 2S Lo-Vac SEM system was used with $<1\,$nm
resolution at optimum 
working distances, a suite of in-column detectors, and two retractable back-scatter detectors.
The images were captured with a resolution of 135 nm at $1000x$ magnification, accelerating voltage of 5 kV, beam current of 25 nA, and working distance of about $10$ mm to highlight the evolution of both composition and topography of sandstone specimens due to mineralization developed during Series~II and Series~III.

Secondary electron (SE) imaging emphasizes surface topography, yielding detailed visualizations of shapes and edges. Conversely, back-scattered electron (BSE) imaging generates contrast based on atomic number, with higher atomic number elements appearing brighter, thus facilitating identification of material composition and phases.

\section{Results} \label{sec:results}

\subsection{SLDV testing} \label{sub:LDV} 

\begin{figure}[]
\center
\includegraphics[width = 0.85\textwidth]{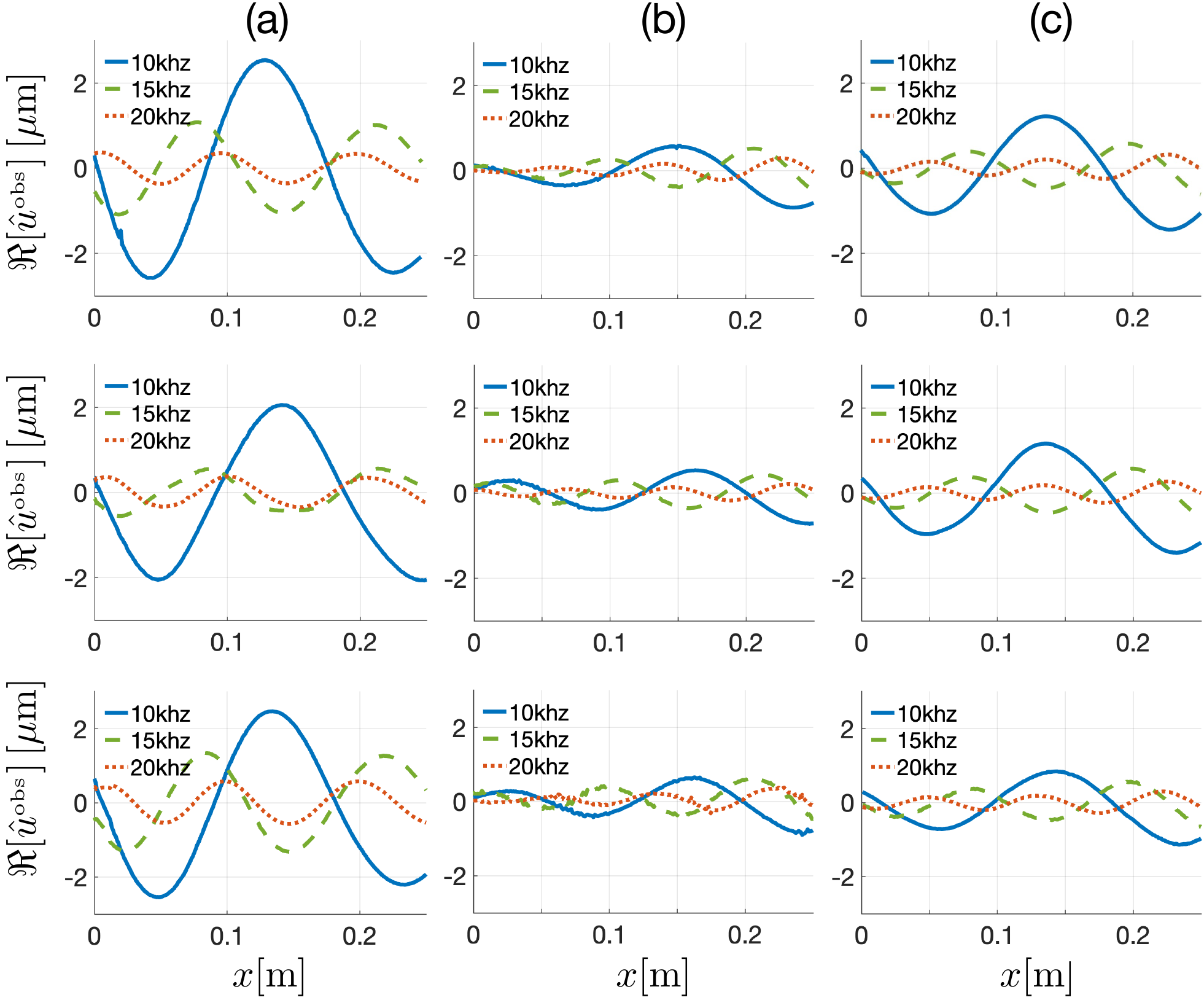}
\caption{Real part of the ultrasonic displacement wavefields $\hat{u}{}^{\obs}(x,\omega=2\pi\hh f_c)$ captured by SLDV at center frequencies $f_c\in\{10,15,20\}\,$kHz  of the excitation wavelet during (a) Series~I testing, (b) Series~II testing, and (c) Series~III testing on: Specimen~I (top row), Specimen~II (middle row) and Specimen~III (bottom row).}
\label{fig:u}
\end{figure}

Three series of SLDV experiments (one “intact” and two “mineralized”) were performed as described in Sec.~\ref{sub:procedure} on the sandstone specimens S1, S2 and~S3. As an illustration of the captured sensory data, Fig.~\ref{fig:laser}(b) shows a snapshot in time of the axial particle velocity wavefield $v^{\obs}(x,t)$. This signal is converted via fast Fourier transform to the frequency domain, $\hat{v}{}^{\obs}(x,\omega)$, yielding the sought displacement wavefield (used by the MECR analysis) as $\hat{u}{}^{\obs} = \ii\hh \hat{v}{}^{\obs}/\omega$. The experimental results verified our 1D wave motion assumption in three distinct ways: (i) ultrasonic displacements in both transverse directions were measured at less than 10\% of their longitudinal counterpart; (ii) longitudinal displacements measured at multiple points along the rod perimeter with the same elevation were practically equal, and (iii) the wavefields $\hat{u}{}^{\obs}(x,\omega)$ captured on the opposing sides of a specimen were found to have negligible variation. Example variations of $\hat{u}{}^{\obs}(x,\omega)$ obtained during the three series of SLDV experiments on specimens~S1 through~S3 are shown in Fig.~\ref{fig:u}. Note that the results are plotted in the direction of reactive flow ($x$ increases bottom-to-top). From the display, one observes that the amplitudes of~$\hat{u}{}^{\obs}$ show a considerable decrease from Series~I (intact rock) to Series~II (mineralized rock), and a modest increase from Series~II to Series~III (re-mineralized rock). 

\subsection{MECR reconstruction} \label{sub:ECR} 

Having obtained the multi-frequency ultrasonic measurements $\hat{u}{}^{\obs}(x,\omega_m)$ with
\[
\omega_m = 10\pi\hh (1+m)\times 10^3\,\text{rad/s}, \qquad m=\overline{1,3}, 
\]
our first objective is to reconstruct the vector of viscoelastic parameters $\bp\!=\!(C,\eta)$ characterizing the sandstone rod as described in Sec.~\ref{MECR-method}, where $C$ and~$\eta$ denote respectively the elastic and viscous modulus of the Maxwell model shown in Fig.~\ref{fig:viscomodels}(a). To expose the heterogeneity of either intact or reacted specimens, we describe the spatial variations of $C(x)$ and $\eta(x)$ by ``pixelized'', i.e. \emph{piecewise-constant} distributions with the pixel size set to $h\!=\!12.5\,$mm. The observation window (over which $\hat{u}{}^{\obs}$ are captured by SLDV measurements) is given by   
\[
\Omega = \{x: 0<x<250\,\text{mm}\}, 
\]
where $x\!=\!0$ is located 40$\,$mm from the bottom of the specimen. This region is then subdivided into three \emph{overlapping subzones} $\Omega_j$ ($j\!=\!\overline{1,3}$) -- each 16 pixels (200$\,$mm) long, with 14 pixels overlap. With reference to the analysis in Sec.~\ref{MECR-method}, the MECR inversion is then performed independently over each subzone~$\Omega_j$ featuring the vector of model parameters $\bp=(C_1,\eta_1,\ldots,C_{16},\eta_{16})$ with two unknowns per pixel. As examined earlier, the nonlinear minimization of~\eqref{mecr3} with~$\Lambda_\kappa$ given by~\eqref{mecr5} entails repeated solution of the coupled forward-backward problem (over~$\Omega_j$ in this case) that is solved via NGSolve -- an open source, Python-based finite element platform~\cite{NGSolve}. A 1D mesh with 100 finite elements of order $p\!=\!3$ is used, and the experimental data $\hat{u}{}^{\obs}$ is mapped onto this mesh by the VoxelCoefficient algorithm available through NGSolve. Starting from a uniform initial guess, the non-quadratic minimization is performed via the sequential least square programming (SLSQP) algorithm available through the Python Scipy library~\cite{scipy2020}. 

\begin{figure}[]
\center
\includegraphics[width=0.95\textwidth]{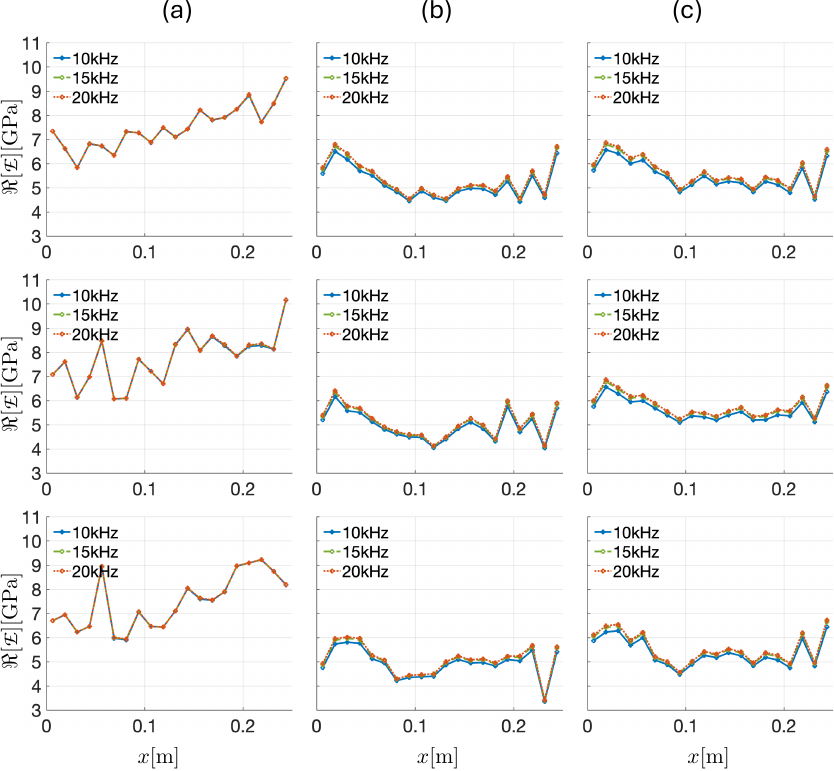}
\caption{Reconstructed variation of the complex Young's modulus $\mathpzc{E}(x,\omega)$, real part: (a) Series~I, (b) Series~II, and (c) Series~III testing performed on specimen S1 (top row), S2 (middle row) and S3 (botom row). Axial coordinate~$x$ increases in the direction of reactive fluid transport.}
\label{fig:E}
\end{figure}

\begin{figure}[]
\center
\includegraphics[width=0.95\textwidth]{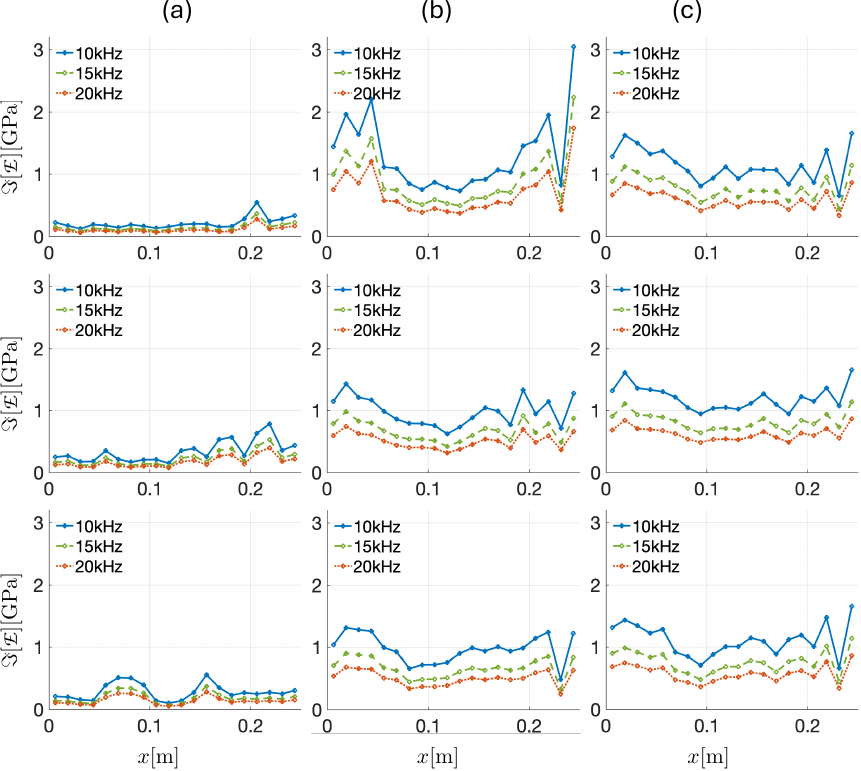}
\caption{Reconstructed variation of the complex Young's modulus $\mathpzc{E}(x,\omega)$, imaginary part: (a) Series~I, (b) Series~II, and (c) Series~III testing performed on specimen S1 (top row), S2 (middle row) and S3 (botom row). Axial coordinate~$x$ increases in the direction of reactive fluid transport.}
\label{fig:alpha}
\end{figure}

On averaging the results of the three MECR reconstructions (over~$\Omega_j$) over common pixels, we obtain piecewise-constant reconstructions~$C^\star(x)$ and~$\eta^\star(x)$ giving the complex Young's modulus  
\[
\mathpzc{E}(x,\omega) = \frac{C^\star(x)\,\eta^\star(x)\, \omega}{\ii\hh C^\star(x) + \eta^\star(x)\,\omega} 
\]
due to~\eqref{mod:complex}. The results of the above computations are illustrated in 
Fig.~\ref{fig:E} and Fig.~\ref{fig:alpha} which plot the reconstructed profiles $\Re[\mathpzc{E}]$ and $\Im[\mathpzc{E}]$, respectively, for each stage of the SLDV testing and all specimens tested (S1 through S3). As seen from Fig.~\ref{fig:E}, the value of  $\Re[\mathpzc{E}]$ drops from about 7.5$\,$GPa in Series~I to about 5.5$\,$GPa in Series~II, and then stays roughly unchanged going to Series~III. On a more granular level, one observes that $\Re[\mathpzc{E}]$ in Series~I \emph{increases with~$x$} for all three (intact) specimens. This trend is \emph{completely reversed} in Series~II and Series~III testing performed on mineralized specimens, possibly because of the evolving geometry of the evaporation front with~$x$. In a similar mineralization setup applied to rod-shaped limestone specimens~\cite{Noiriel2010}, the evaporation front was observed to become increasingly parabolic away from the feed point. In the context of our study, this suggests an increasing degree of mineralization -- and so damage -- with~$x$ due to growing area of the evaporation front. For completeness, Fig.~\ref{fig:alpha} plots the variation of $\Im[\mathpzc{E}]$ for the three series of tests performed on specimens S1 through S3. In Series~I, $\Im[\mathpzc{E}]$ is found to be relatively low, showing the value of about 0.15$\,$GPa at 15kHz for all three intact specimens. The Series~II results, however, show a major jump in $\Im[\mathpzc{E}]$ to roughly 0.6$\,$GPa at 15kHz. By contrast, there is little difference in the reconstructed profiles of  $\Im[\mathpzc{E}]$ between Series~II and Series~III. 

\begin{figure}[]
\center
\includegraphics[width=0.95\textwidth]{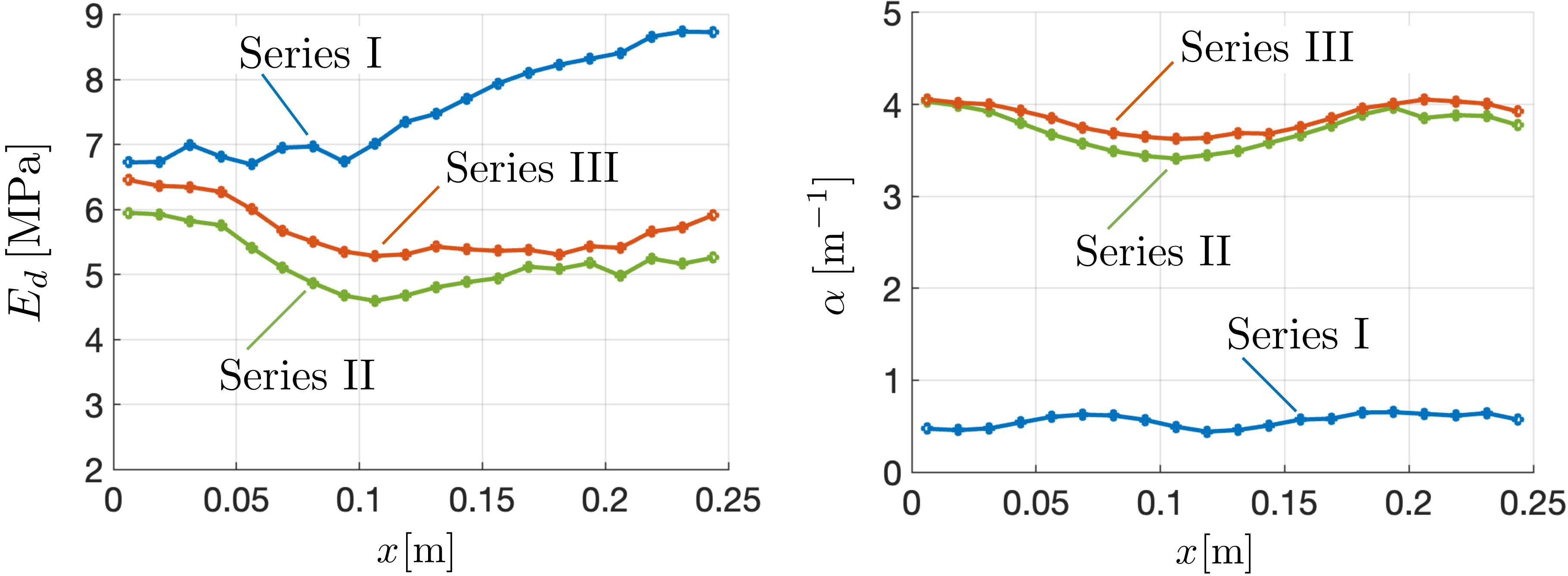}
\caption{Reconstructed profiles of the dynamic Young's modulus~$E_d$ and attenuation coefficient~$\alpha$ at 10$\,$kHz: pointwise averaged values for the specimens S1 through S3.}
\label{sldv-final}
\end{figure}

On the basis of the above results, the (frequency-dependent) dynamic Young' modulus and attenuation of the sandstone specimens are computed respectively as  
\begin{equation}\label{Edalfax}
E_d(x,\omega)=\Re[\mathpzc{E}(x,\omega)], \qquad
\alpha(x,\omega) = \Im\Big[ \frac{\omega}{\sqrt{\mathpzc{E}(x,\omega) / \rho}} \Big] 
\end{equation}
according to~\eqref{Edalfa}. The profiles of~$E_d(x,\omega_1)$ and $\alpha(x,\omega_1)$, averaged pointwise over the three specimens (S1 through S3), are plotted in Fig.~\ref{sldv-final}. Comparing the 10$\,$kHz ultrasonic properties of mineralized specimens  (Series~II and Series~III testing) to their intact values (Series~I testing), we observe (i)~about 20-25\% decrease in dynamic modulus that grows with distance from the feed point, and (ii)~roughly 7-fold increase in the attenuation coefficient. Here it is worth noting that the dynamic Young's modulus of intact Dunville sandstone (Series~I), $E_d\simeq 7.5\,$MPa, is larger than its nominal value $E=6.2\,$MPa listed in Table~\ref{tab:properties}. Given the fact that $E_d$ in Fig.~\ref{sldv-final} is reconstructed at 10$\,$kHz while $E$ is measured at 1$\,$MHz, this difference is consistent with the experimental observations of \emph{negative dispersion} in high-porosity sandstones \cite{winkler1983frequency} at frequencies between 400$\,$kHz and 2$\,$MHz. 

\subsection{Acoustic emission (AE) monitoring} \label{sub:AE} 

\begin{figure}[H]
\center
\includegraphics[width=0.4\textwidth]{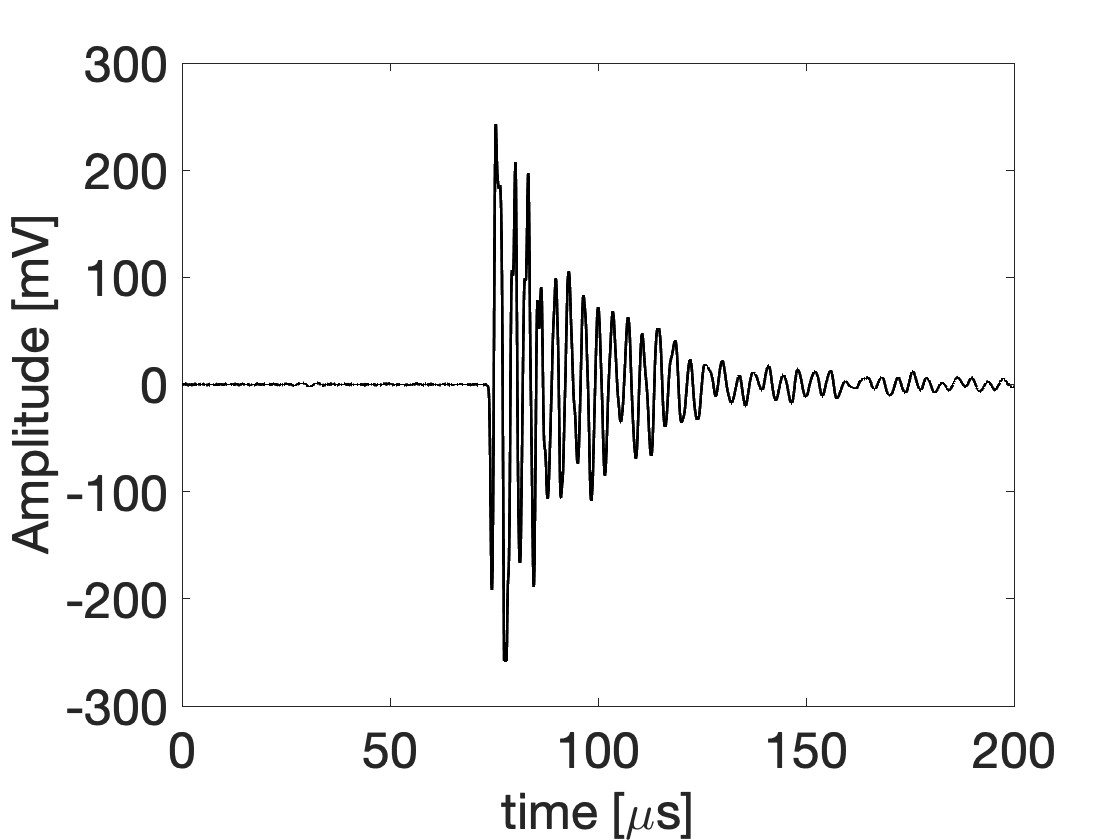} 
\caption{Typical trace of an AE record recorded during mineralization.}
\label{fig:ae_pattern}
\end{figure}

No AE events were observed during the first five days of the initial (two-week) mineralization period. Afterwards, the events began to emerge and their frequency increased steadily with time. Nonetheless, only 20~events were recorded during the 14 days of mineralization. Fig.~\ref{fig:ae_pattern} shows a representative sample of recorded events with dominant frequencies within the range of 200-400$\,$kHz, falling well inside the AS sensor bandwidth of 125-750$\,$kHz. The sparsity of the events suggests that limited microcracking occurred during the mineralization process, which is consistent with the ``unconfined'' experimental setup  where the reactive fluid was able to easily exit the specimen and mineralize on its outer surface, see for instance Fig.~\ref{fig:specimens}. Nonetheless, the combination of such (apparently limited) damage and mineral deposition resulted in a clearly observable decrease in dynamic modulus and a major increase in the attenuation coefficient, highlighting the potential of ultrasonic (and so seismic) waves for tracking the progress of mineralization in rock masses.
  
\subsection{X-ray tomography} \label{sub:CT} 

The $\mu$CT interrogation of Dunnville sandstone was primarily focused on the porosity and micro-scale characteristics of both intact and mineralized samples. Tomography scans were performed on 20$\,$mm-long, 5$\,$mm-diameter cylindrical samples extracted around the midpoint of each section examined (top, middle, and bottom) of specimens S5 and S4 to assess the variations in sandstone porosity before mineralization (representative of Series~I testing) and after mineralization (representative of Series~II testing), respectively. The $\mu$CT scans underwent noise reduction filtering  to suppress unwanted random variations, followed by Gaussian filtering for enhanced edge detection and segmentation. The sample images were subsequently segmented into matrix and porosity constituents, allowing for the visualization of the rock's microstructure and its constituents.

\begin{figure}[H]
\center
\includegraphics[width=0.75\textwidth]{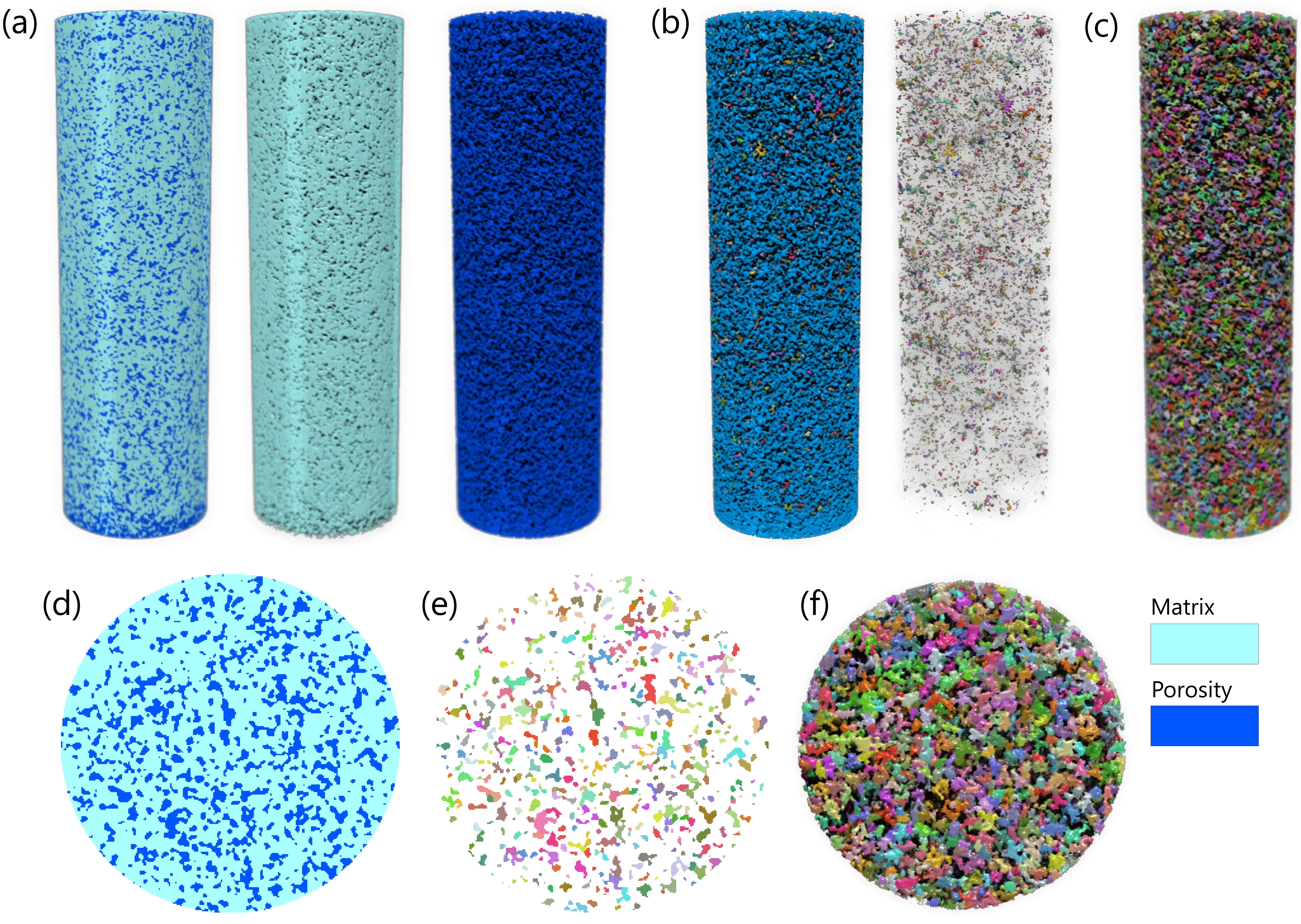}   
\caption{Segmentation and microstructural analysis of 3D volume (top row) and  mid-height cross section of the intact S5 specimen (bottom section): (a) and (d) -- segmentation of sample into the matrix and porosity constituents, presented from left to right for the entire rock, matrix, and porosity; (b) and~(e) -- multi-ROI analysis of the extracted porosity, with (in blue color) and without the largest connected element; (c) and~(f) pore network extraction of the largest connected multi-ROI pore volume, color-coded to distinguish separate pore bodies.}
\label{fig:Ftomo}
\end{figure}

Panels~(a) and~(d) in Fig.~\ref{fig:Ftomo} depict respectively the segmented 3D volume and mid-height cross section of a cylindrical sample taken from the intact specimen S5 (bottom section, see Fig.~\ref{fig:specimens}(c)), with the matrix and pore space color-coded for better visualization. Analysis of the $\mu$CT data revealed an average porosity of 26.8\% \(\pm\) 0.8\%, consistently observed across all samples. Both intact and mineralized specimens exhibited similar standard deviations of approximately 1\% across different sections.

The segmented porosity volume, representing the void spaces within the specimen, was then extracted for further analysis of pore connectivity and distribution. A multi-region of interest (multi-ROI) analysis was performed on the segmented porosity volume, which served as input to the connected-component labeling algorithm. Multi-ROI analysis involves identifying and isolating distinct connected components within the pore space, where a connected component is defined as a set of adjacent voxels sharing the same phase (i.e., porosity) based on connectivity in 3D space. This process enabled the identification of the largest connected element, the most extensive continuous pore network within the sample, as well as smaller, localized connected pore bodies, which are isolated pore clusters disconnected from the main network. The results, shown in panels~(b) and~(e) of  Fig.~\ref{fig:Ftomo}, depict the multi-ROI analysis of the extracted porosity, with the largest connected element highlighted in blue color and other pore bodies color-coded individually to distinguish their spatial distribution. Specifically, the left image in panel~(b) include all connected pore bodies, while the right panel excludes the largest connected element to emphasize smaller, localized pores.

The largest connected multi-ROI was then used as the input for pore network extraction using the OpenPNM package \cite{gostick2016openpnm}. The largest connected multi-ROI pore volume was processed to generate a pore-throat network, where pore bodies are defined as larger void regions and throats as narrower constrictions connecting them. The extraction algorithm identified individual pore bodies and throats by approximating the pore space as a network of spherical pores connected by cylindrical throats. Each pore body was assigned an equivalent pore diameter, calculated as the diameter of a sphere with the same volume as the pore, and a pore connectivity index, representing the number of throats connecting each pore to its neighbors. The resulting individual pore network map is visualized in panels~(c) and~(f) of Fig.~\ref{fig:Ftomo}, depicting respectively the 3D volume and mid-height cross section, with the individual pore bodies color-coded to distinguish their spatial distribution. Analysis of the pore-throat computations and maps revealed equivalent pore diameters varying within 120-230$\,\mu$m and pore connectivity indices varying between 2 and 9, indicating moderate to high connectivity of the pore network.

Analysis of the $\mu$CT data at 20$\,\mu$m voxel resolution revealed no significant differences in porosity between the intact S5 specimen (Series~I) and reacted S4 specimen (Series~II). Pore network modeling showed that micro-scale pore-throat characteristics, including equivalent pore diameters and connectivity indices, were statistically similar for both specimens. These findings suggest that halite crystallization in the reacted S4 specimen did not significantly alter the pore geometry or connectivity at the observed spatial scale.

\begin{figure}[]
\center
\includegraphics[height=4 cm]{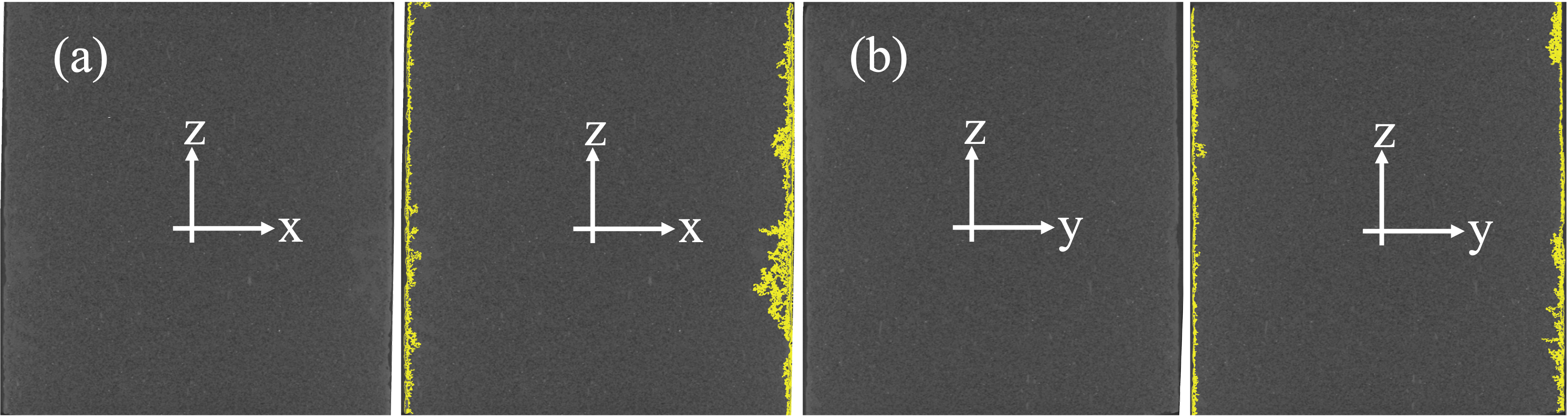}   
\caption{Raw (left panels) and segmented (right panels) images of the ``vertical'' slices taken from the bottom section of re-mineralized specimen S3, corresponding to Series~III testing: (a) $x-z$ plane and (b) $y-z$ plane, where $z$ is the axis of symmetry of the sample.}
\label{fig:SEM}
\end{figure}

Fig.~\ref{fig:SEM} shows the raw and segmented tomography images of vertical cross sections obtained from the bottom section of re-mineralized S3 specimen (Series~III). The images reveal halite accumulations (color-coded in yellow) concentrated near the specimen’s outer boundary. Halite accumulations were segmented using grey-value tracking, targeting surface and near-surface regions with established spatial continuity. The surface and near-surface halite concentration indicates a preference for efflorescence (surface crystallization) over subflorescence (internal crystallization), consistent with the reported precipitation patterns of sodium chloride in porous media ~\cite{veran2014evaporation, derluyn2024experimental}.

\subsection{Scanning electron microscopy} \label{sub:SEM2} 

\begin{figure}[h]
\center
\includegraphics[width = 0.95\textwidth]{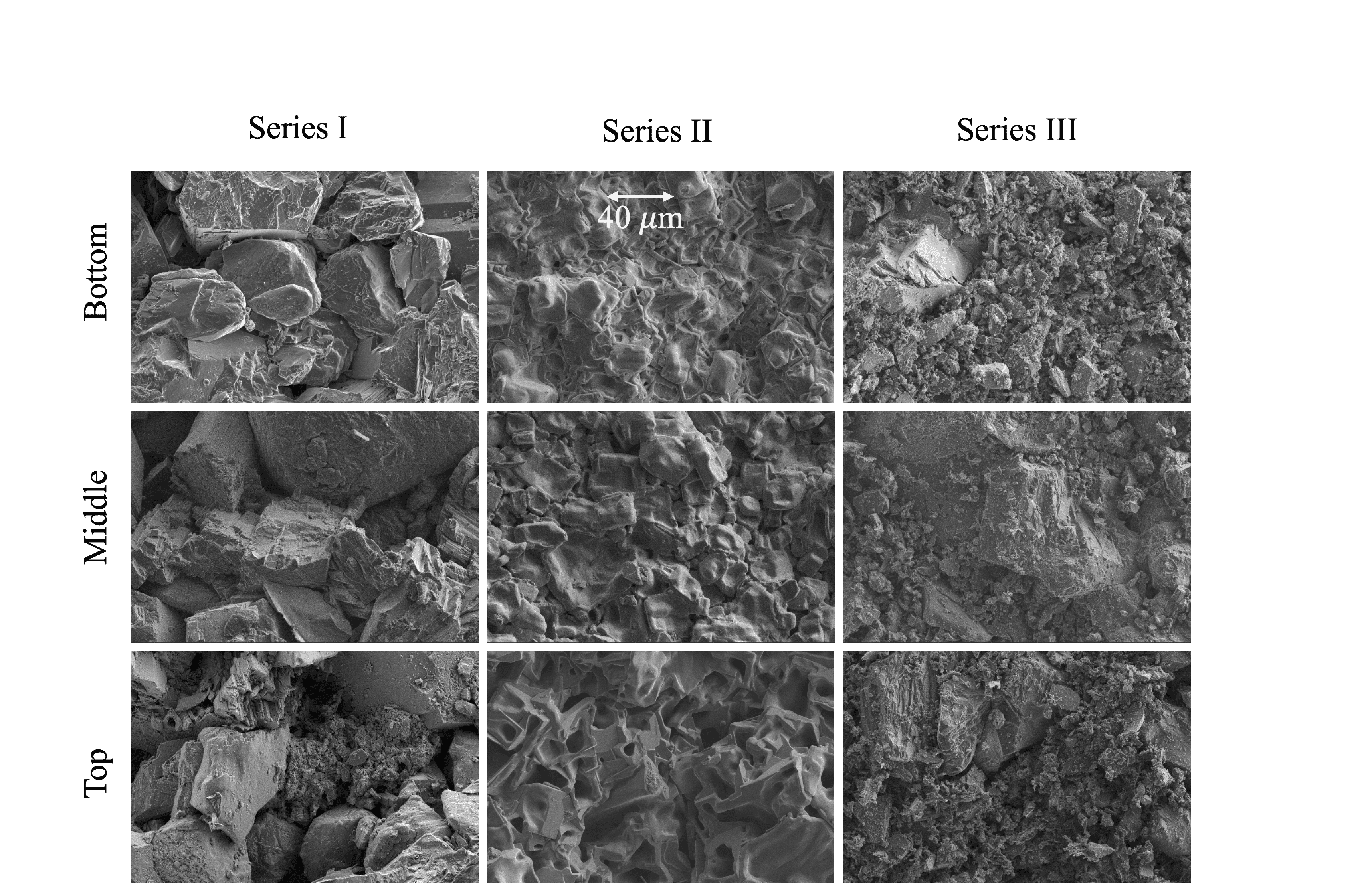}
\caption{SE microscopy images (at 135$\,$nm resolution) of the bottom, middle and top sections of: specimen~S5 (Series~I), specimen~S4 (Series~II), and specimen~S3 (Series~III).}
\label{fig:I_II_III_topography}
\end{figure}

\begin{figure}[h]
\center
\includegraphics[width = 0.95\textwidth]{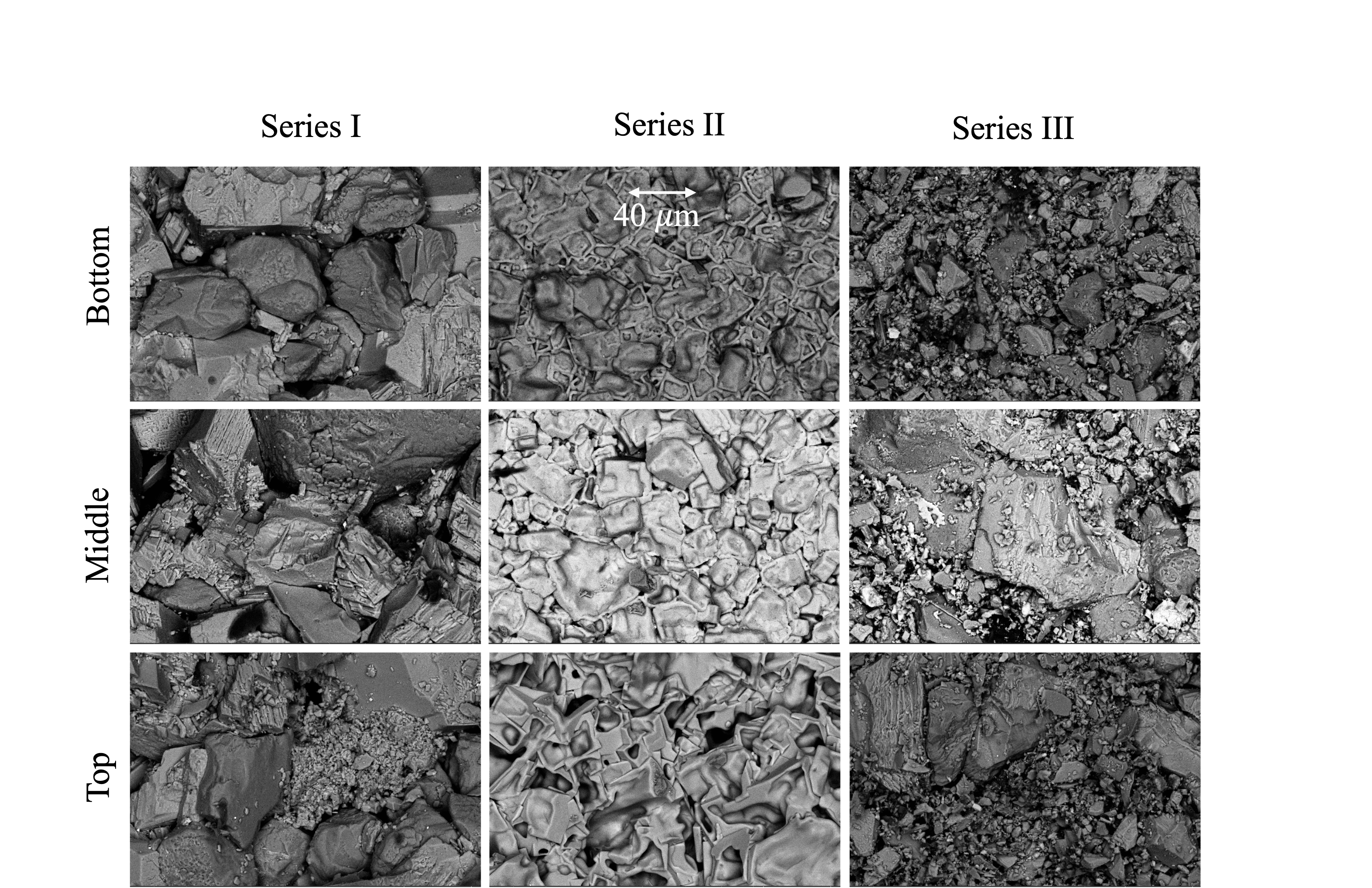}
\caption{Composition obtained by BSE microscopy at 135$\,$nm  resolution for the bottom, middle and top sections of: specimen~S5 (Series~I), specimen~S4 (Series~II), and specimen~S3 (Series~III).}
\label{fig:I_II_III_composition}
\end{figure}

\begin{figure}[h]
\center
\includegraphics[width = 0.65\textwidth]{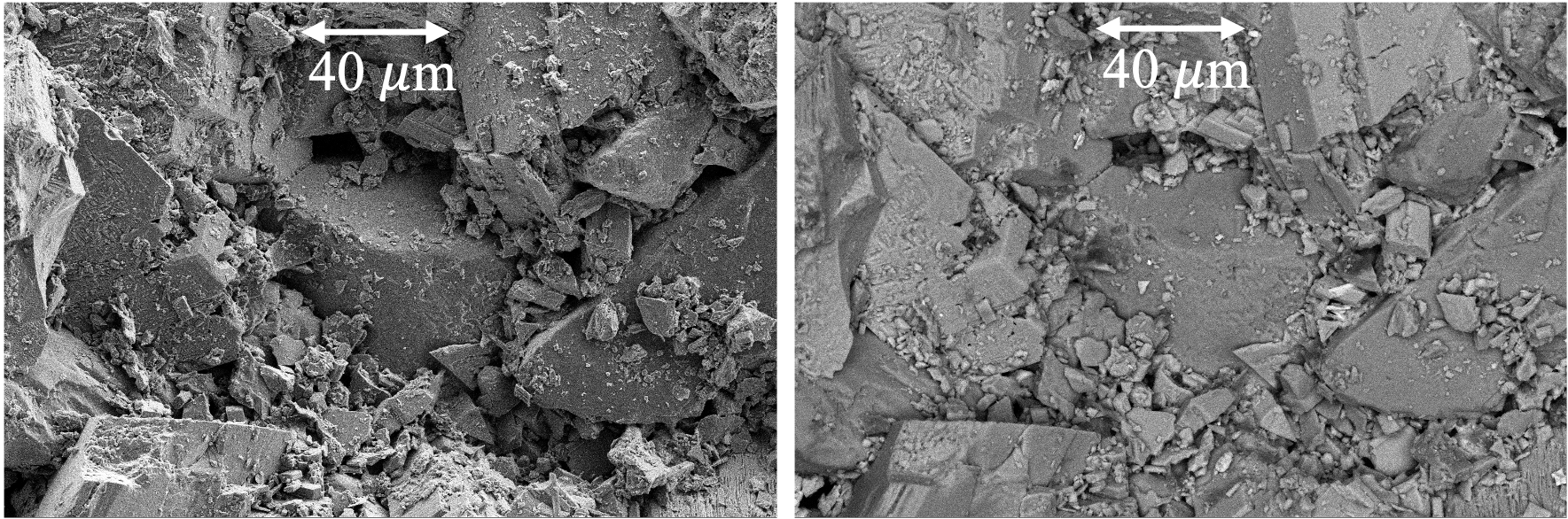}
\caption{Bottom section of Series~II washed with water and its upper face scanned with a resolution of 135 nm using secondary (left) and back-scattered (right) electrons.}
\label{fig:II_washed}
\end{figure}

Fig.~\ref{fig:I_II_III_topography} and Fig.~\ref{fig:I_II_III_composition} show respectively the scanning electron (SE) and backscattered scanning electron (BSE) microscopy images taken from the bottom, middle, and top sections of an intact rod (specimen~S5, Series~I), mineralized rod (specimen~S4, Series~II), and  re-mineralized rod (specimen~S3, Series~III). As can be seen from the displays, mineralized sections of specimen~S4 are completely covered with halite precipitates, whereas re-mineralized sections of specimen~S3 feature significant accumulation of powder-like precipitate matter -- possibly commingled with spalled-off grain fragments. To help expose the microstructure of ``coated'' mineralized sections, the bottom section of specimen~S4 was washed with deionized (DI) water and re-scanned by SE and BSE miroscopy, with the results shown in Fig.~\ref{fig:II_washed}. Overall, the supplementary images show a structure  with significant accumulation of powder-like precipitate matter -- resembling that of specimen~S3 in Fig.~\ref{fig:I_II_III_topography} and Fig.~\ref{fig:I_II_III_composition}.


\begin{figure}[h]
\center
\includegraphics[width = 0.8\textwidth]{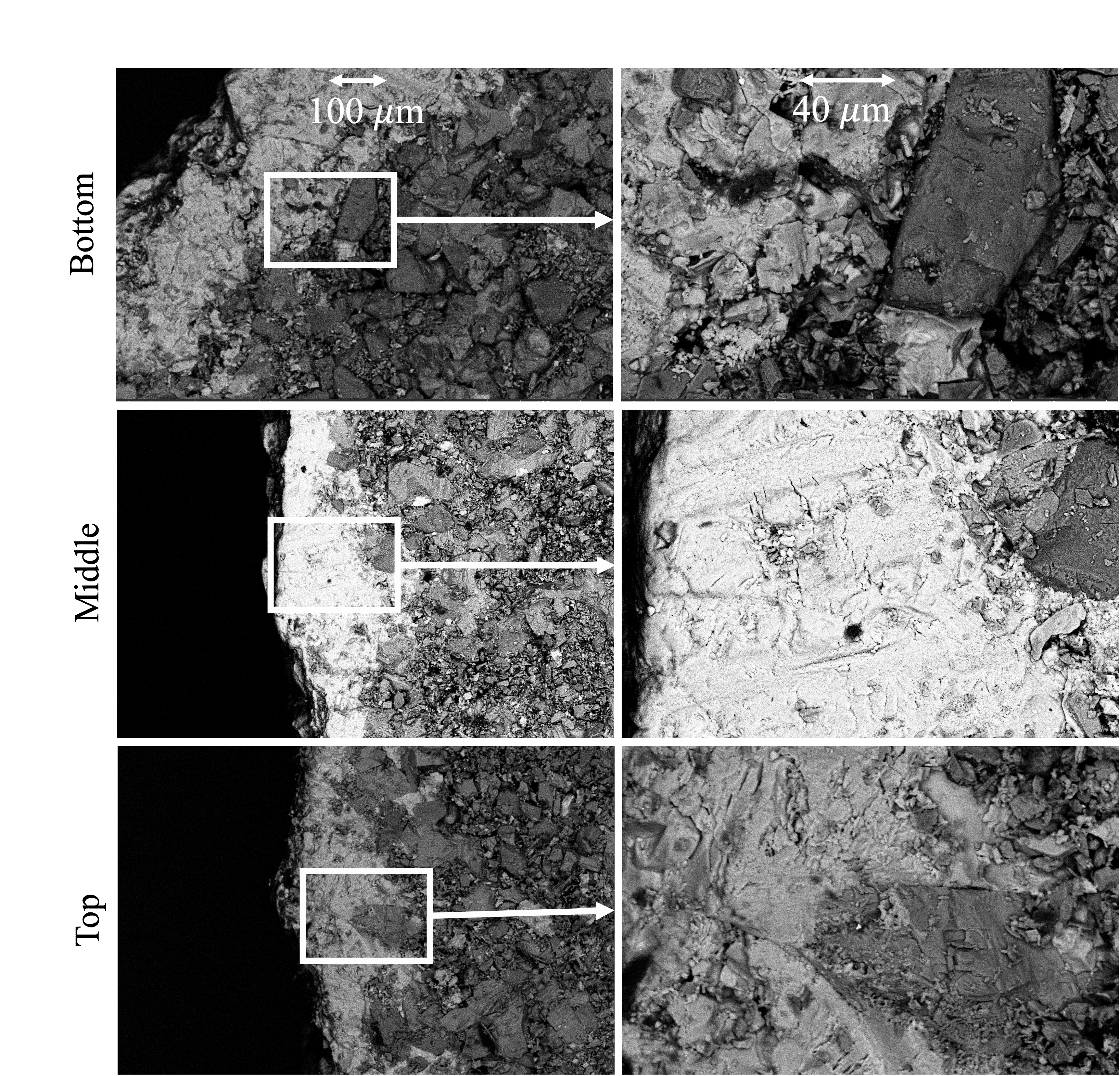}
\caption{Near-perimeter scans of the bottom, middle, and top sections of specimen~S3 (Series~III) obtained by BSE microscopy with a resolution of 539 nm (left column) and 135 nm (right column).}
\label{fig:I_II_III_surface}
\end{figure}

For completeness, we also scanned by BSE microscopy the perimeter of the featured nine sections. For Series~I and Series~II, there is not much change near the outer surface (results not shown); however for Series~III, a significant amount of salt is seen to have mineralized near the perimeter of all three sections (bottom, middle, and top) as shown in Fig.~\ref{fig:I_II_III_surface}.

\section{Discussion} \label{sec:discussion}

\subsection{Mechanical properties of intact and mineralized specimens} \label{sub:intactvsmineralized} 

As can be seen from Fig.~\ref{fig:E} and the left panel of Fig.~\ref{sldv-final}, dynamic Young's modulus $E_d$ of intact specimens (Series~I) exhibits a distinct increase in  direction~$x$ of the reactive transport (perpendicular to the bedding plane), a feature that is common to specimens S1 through S3 and can be attributed to \emph{in-situ} variations of Dunnville sandstone. The attenuation coefficient~$\alpha$ of intact rock also shows some natural fluctuations, albeit without a clear trend. The results for reacted specimens, on the other hand, show an opposite trend where~$E_d$ decreases with~$x$. For Series~II (resp. Series~III), we observe from Fig.~\ref{sldv-final} that the ``damage ratio'' $\Delta E_d/E_d^{intact}$ increases almost monotonically from 0.098 (resp. 0.062) at $x\!=\!0$ to 0.357 (resp. 0.332) at  $x\!\simeq\!25\,$cm. By contrast, the ``dissipation ratio'' $\Delta\alpha/\alpha^{intact}$ is (i)~much more pronounced with the respective mean values of 5.8 and 5.9 for Series~II and Series~III, and (ii) void of clear trends in terms of spatial variation versus~$x$. 

The systemic decrease in dynamic Young's modulus from Series~I to Series~II testing observed in Fig.~\ref{sldv-final} can be attributed to mineralization-driven microcracking, while the slight increase in~$E_d$ from Series~II to Series~III testing is likely the result of additional pore filling due to re-mineralization reflected in the SEM images in Fig.~\ref{fig:I_II_III_topography} and Fig.~\ref{fig:I_II_III_composition}. Here it us useful to recall that neither microcracking nor pore filling are generally uniform in the radial direction, whereby the reconstructed $E_d$ profiles should be understood as those of an \emph{effective} dynamic Young's modulus, ``homogenized'' across each cross section. At a more granular level, it could be argued with reference to Fig. \ref{fig:I_II_III_surface} that the obtained $E_d$ maps are the result of two \emph{mutually-counteracting} trends, namely the pore filling in the near-perimeter region and (limited) microcracking in the interior region.

Perhaps more strikingly, the results from Fig.~\ref{fig:alpha} and the right panel in Fig.~\ref{sldv-final} demonstrate a \emph{7-fold increase} in the attenuation coefficient~$\alpha$ of mineralized specimens (Series~II and Series~III) relative to that of the intact specimens (Series~I). This phenomenon is likely the result of two \emph{mutually-reinforcing} trends, namely microcracking and pore filling with powder-like precipitate, both of which contribute to increased friction and so dissipation in the the system. While the friction by itself generates rate-independent dissipation, here it must be recognized that effective ultrasonic (or seismic) attenuation is the cumulative result of (i) local friction and (ii) wave scattering at the microscale that is wavelength -- and so frequency -- dependent. In the experiments performed in this study, these two factors result in the attenuation that grows approximately linearly with frequency (in a ``Maxwell-like'' fashion) between 10$\,$kHz and 20$\,$kHz.

\subsection{Veracity of $\mu$CT and SEM scans} \label{sub:CT_discussion} 

From the results presented in Sec.~\ref{sub:CT}, it is apparent that the X-ray tomography with $O(\mu\text{m})$ voxel resolution does not show a clear difference between intact and mineralized specimens. Instead, as seen from Fig.~\ref{fig:I_II_III_surface}, it helped expose the accumulation of halite minerals near the perimeter of re-mineralized specimens (Series~III testing). These results suggest that the micro-structural changes due to mineralization are on the  $o(\mu\text{m})$ scale. To mitigate the impediment, we performed additional SEM scans with $O(\text{nm})$ resolution, which exposed a powder-like precipitate matter in the pore space of (re-) mineralized specimens -- possibly commingled with spalled-off grain fragments. As examined earlier, the MECR-based ultrasonic interrogation exposes the macroscopic (mechanical) effect of these microscopic alterations, with an additional benefit of revealing the spatial variation of such changes.

\section{Summary} \label{sec:conclusions}

In this study, we investigate the (macroscopic) mechanical imprint of reactive fluid transport in porous rock via ultrasonic interrogation. To this end, we expose rod-like specimens of Dunnville sandstone to NaCl transport (by capillary action) and mineralization (by evaporation). To understand the mechanical effect of reactive transport on a macroscopic scale, we pursue ultrasonic characterization of rod-like specimens by way of scanning laser Doppler vibrometry and modified-error-in-constitutive-relation (MECR) back-analysis of the full-field ultrasonic data (i)~before mineralization, (ii) after two weeks of mineralization, and (iii) after four weeks of mineralization. Our results, computed by averaging the results obtained from three (identically prepared and reacted) specimens demonstrate (a) roughly 30\% drop in Young's modulus, and (b) 7-fold increase in attenuation coefficient due to mineralization. On a more granular level, the MECR reconstruction revealed that the damage of Young's modulus increases with distance from the ``feed point'' (the end of the rod submerged in reactive fluid), while the increase in attenuation is uniform throughout the specimen. To better understand the root causes of these changes, we made use of the X-ray micro-computed tomography ($\mu$CT) and scanning electron microscopy (SEM) of selected cross-sections. The grain-scale information suggests that microcracking, accompanied by pore filling with powder-like precipitate, is  responsible for the observed macroscopic changes.

\section*{Acknowledgements}

\noindent This work was supported as part of the \emph{Center on Geo-Processes in Mineral Carbon Storage}, an Energy Frontier Research Center funded by the U.S. Department of Energy, Office of Science, Basic Energy Sciences, at the University of Minnesota under award \# DE-SC0023429.

\bibliographystyle{unsrt}
\bibliography{fingerprint1D}

\begin{thebibliography}{10}

\bibitem{Chhabra2022}
R.~Chhabra.
\newblock {\em Salt-affected Soils and Marginal Waters: Global Perspectives and
  Sustainable Management}.
\newblock Springer Nature, 2022.

\bibitem{Hall2001}
K.~Hall and M-F. Andr{\'e}.
\newblock New insights into rock weathering from high-frequency rock
  temperature data: an {A}ntarctic study of weathering by thermal stress.
\newblock {\em Geomorphology}, 41:23--35, 2001.

\bibitem{Pel2003}
L.~Pel, H.~Huinink, and K.~Kopinga.
\newblock Salt transport and crystallization in porous building materials.
\newblock {\em Magnetic Resonance Imaging}, 21:317--320, 2003.

\bibitem{Mi2020}
X.~Mi, T.~Li, J.~Wang, and Y.~Hu.
\newblock Evaluation of salt-induced damage to aged wood of historical wooden
  buildings.
\newblock {\em International Journal of Analytical Chemistry}, 2020:8873713,
  2020.

\bibitem{Taber1916}
S.W. Taber.
\newblock The growth of crystals under external pressure.
\newblock {\em American Journal of Science}, 41:532--556, 1916.

\bibitem{Steiger2005}
M.~Steiger.
\newblock Crystal growth in porous materials - {II}: Influence of crystal size
  on the crystallization pressure.
\newblock {\em Journal of Crystal Growth}, 282:470--481, 2005.

\bibitem{Doehne2010}
C.A. Price and E.~Doehne.
\newblock {\em {Stone Conservation: an Overview of Current Research}}.
\newblock Getty Publications, 2011.

\bibitem{shahidzadeh2024crystallization}
N.~Shahidzadeh.
\newblock Crystallization pressure.
\newblock In {\em Salt Crystallization in Porous Media}, pages 25--44. Wiley,
  2024.

\bibitem{Khodabandeh2022}
M.A. Khodabandeh and N.~Rozgonyi-Boissinot.
\newblock The effect of salt weathering and water absorption on the ultrasonic
  pulse velocities of highly porous limestone.
\newblock {\em Periodica Polytechnica Civil Engineering}, 66:627--639, 2022.

\bibitem{nooraiepour2025potential}
M.~Nooraiepour, K.~Pola{\'n}ski, and M.~Masoudi et~al.
\newblock Potential for 50\% mechanical strength decline in sandstone
  reservoirs due to salt precipitation and {CO2}--brine interactions during
  carbon sequestration.
\newblock {\em Rock Mechanics and Rock Engineering}, 58:1239--1269, 2025.

\bibitem{Ringrose2019}
P.S. Ringrose and T.A. Meckel.
\newblock Maturing global {CO2} storage resources on offshore continental
  margins to achieve {2DS} emissions reductions.
\newblock {\em Scientific Reports}, 9, 2019.

\bibitem{Bruant2002}
R.G. Bruant, A.J. Guswa, M.A. Celia, and C.A. Peters.
\newblock Safe storage of {CO2} in deep saline aquifiers.
\newblock {\em Environmental Science \& Technology}, 36:240A--245A, 2002.

\bibitem{Espinoza2018}
D.N. Espinoza, H.~Jung, J.R. Major, Z.~Sun, M.J. Ramos, P.~Eichhubl, M.T.
  Balhoff, R.C. Choens, and T.A. Dewers.
\newblock {CO2} charged brines changed rock strength and stiffness at {C}rystal
  {G}eyser, {U}tah: Implications for leaking subsurface {CO2} storage
  reservoirs.
\newblock {\em International Journal of Greenhouse Gas Control}, 73:16--28,
  2018.

\bibitem{Rathnaweera2015}
T.D. Rathnaweera, P.G. Ranjith, and M.S.A. et~al. Perera.
\newblock {CO2-induced mechanical behaviour of Hawkesbury sandstone in the
  Gosford basin: An experimental study}.
\newblock {\em Materials Science and Engineering: A}, 641:123--137, 2015.

\bibitem{Masoudi2021}
M.~Masoudi, H.~Fazeli, Miri R., and H.~Hellevang.
\newblock Pore scale modeling and evaluation of clogging behavior of salt
  crystal aggregates in {CO2}-rich phase during carbon storage.
\newblock {\em International Journal of Greenhouse Gas Control}, 111:103475,
  2021.

\bibitem{nooraiepour2018effect}
M.~Nooraiepour, H.~Fazeli, R.~Miri, and H.~Hellevang.
\newblock Effect of {CO2} phase states and flow rate on salt precipitation in
  shale caprocks—a microfluidic study.
\newblock {\em Environmental Science \& Technology}, 52:6050--6060, 2018.

\bibitem{dkabrowski2025surface}
K.M. Dabrowski, M.~Nooraiepour, M.~Masoudi, M.~Zajac, S.~Kuczynski, R.~Smulski,
  J.~Barbacki, H.~Hellevang, and S.~Nagy.
\newblock Surface wettability governs brine evaporation and salt precipitation
  during carbon sequestration in saline aquifers: Microfluidic insights.
\newblock {\em Science of the Total Environment}, 958:178110, 2025.

\bibitem{Rodriguez1999}
C.~Rodriguez and E.~Doehne.
\newblock Salt weathering: Influence of evaporation rate, supersaturation and
  crystallization pattern.
\newblock {\em Earth Surface Processes and Landforms}, 24:191--209, 1999.

\bibitem{greenleaf2003}
J.F. Greenleaf, M.~Fatemi, and M.~Insana.
\newblock Selected methods for imaging elastic properties of biological
  tissues.
\newblock {\em Annual Review of Biomedical Engineering}, 5:57--78, 2003.

\bibitem{parker2005}
K.J. Parker, L.S. Taylor, S.~Gracewski, and D.J. Rubens.
\newblock A unified view of imaging the elastic properties of tissue.
\newblock {\em The Journal of the Acoustical Society of America},
  117:2705--2712, 2005.

\bibitem{sigrist2017}
R.M.S. Sigrist, J.~Liau, E.K. Ahmed, M.C. Chammas, and J.K. Willmann.
\newblock Ultrasound elastography: review of techniques and clinical
  applications.
\newblock {\em Theranostics}, 7:1303, 2017.

\bibitem{mariappan2010}
Y.K. Mariappan, K.J. Glaser, and R.L. Ehman.
\newblock Magnetic resonance elastography: a review.
\newblock {\em Clinical Anatomy}, 23:497--511, 2010.

\bibitem{Pourahmadian2018}
F.~Pourahmadian and B.B. Guzina.
\newblock On the elastic anatomy of heterogeneous fractures in rock.
\newblock {\em International Journal of Rock Mechanics and Mining Sciences},
  106:259--268, 6 2018.

\bibitem{bonnet2024}
M.~Bonnet, P.~Salasiya, and B.B. Guzina.
\newblock Modified error-in-constitutive-relation ({MECR}) framework for the
  characterization of linear viscoelastic solids.
\newblock {\em Journal of the Mechanics and Physics of Solids}, 190:105746,
  2024.

\bibitem{lad:ned:rey:94}
P.~Ladev{\`e}ze, D.~Nedjar, and M.~Reynier.
\newblock Updating of finite element models using vibration tests.
\newblock {\em AIAA Journal}, 32:1485--1491, 1994.

\bibitem{allix:05}
O.~Allix, P.~Feissel, and H.~M. Nguyen.
\newblock Identification strategy in the presence of corrupted measurements.
\newblock {\em Engineering Computations}, 22:487--504, 2005.

\bibitem{Scherer2004}
G.W. Scherer.
\newblock Stress from crystallization of salt.
\newblock {\em Cement and Concrete Research}, 34:1613--1624, 2004.

\bibitem{Tarokh2022}
A.~Tarokh, J.D. Sharpe, and J.F. Labuz.
\newblock Confined tensile testing of porous sandstone.
\newblock {\em Rock Mechanics and Rock Engineering}, 55:6555--6566, 2022.

\bibitem{guzina2025constitutive}
B.B. Guzina and M.~Bonnet.
\newblock On the constitutive behavior of linear viscoelastic solids under the
  plane stress condition.
\newblock {\em Journal of Elasticity}, 157:1--38, 2025.

\bibitem{diaz2015modified}
M.I. Diaz, W.~Aquino, and M.~Bonnet.
\newblock A modified error in constitutive equation approach for
  frequency-domain viscoelasticity imaging using interior data.
\newblock {\em Computer Methods in Applied Mechanics and Engineering},
  296:129--149, 2015.

\bibitem{banerjee2013}
B.~Banerjee, T.~F. Walsh, W.~Aquino, and M.~Bonnet.
\newblock Large scale parameter estimation problems in frequency-domain
  elastodynamics using an error in constitutive equation functional.
\newblock {\em Computer Methods in Applied Mechanics and Engineering},
  253:60--72, 2013.

\bibitem{nooraiepour2025three}
M.~Nooraiepour, M.~Masoudi, H.~Derluyn, P.~Senechal, P.~Moonen, and
  H.~Hellevang.
\newblock Three-dimensional porous structures of {CO2}-induced salt
  precipitation sustaining halite self-enhancing growth.
\newblock {\em arXiv preprint arXiv:2506.17241}, 2025.

\bibitem{gostick2016openpnm}
J.~Gostick, M~Aghighi, J.~Hinebaugh, T.~Tranter, and M~Hoeh et~al.
\newblock Openpnm: a pore network modeling package.
\newblock {\em Computing in Science \& Engineering}, 18:60--74, 2016.

\bibitem{NGSolve}
J.~Sch{\"o}berl.
\newblock C++ 11 {I}mplementation of {F}inite {E}lements in {NGSolve}.
\newblock {\em Institute for Analysis and Scientific Computing, Vienna
  University of Technology}, 30, 2014.

\bibitem{scipy2020}
P.~Virtanen, R.~Gommers, and T.E. et~al. Oliphant.
\newblock {SciPy} 1.0: fundamental algorithms for scientific computing in
  {P}ython.
\newblock {\em Nature Methods}, 17:261--272, 2020.

\bibitem{Noiriel2010}
C.~Noiriel, F.~Renard, M.L. Doan, and J.P. Gratier.
\newblock Intense fracturing and fracture sealing induced by mineral growth in
  porous rocks.
\newblock {\em Chemical Geology}, 269:197--209, 2010.

\bibitem{winkler1983frequency}
K.W. Winkler.
\newblock Frequency dependent ultrasonic properties of high-porosity
  sandstones.
\newblock {\em Journal of Geophysical Research: Solid Earth}, 88:9493--9499,
  1983.

\bibitem{veran2014evaporation}
S.~Veran-Tissoires and M.~Prat.
\newblock Evaporation of a sodium chloride solution from a saturated porous
  medium with efflorescence formation.
\newblock {\em Journal of Fluid Mechanics}, 749:701--749, 2014.

\bibitem{derluyn2024experimental}
H.~Derluyn.
\newblock Experimental observations on salt crystallization.
\newblock In {\em Salt Crystallization in Porous Media}, pages 99--125. Wiley,
  2024.

\end{thebibliography}

\end{document}